\documentclass[letterpaper, draftcls, 12pt]{IEEEtran}
\usepackage[utf8]{inputenc} 
\usepackage[T1]{fontenc}
\usepackage{url}              
\usepackage{cite}             

\usepackage[cmex10]{amsmath}  
\usepackage{ amssymb }

\DeclareMathOperator*{\argmin}{arg\,min}

\usepackage{mleftright}       
\mleftright                   

\usepackage{graphicx}         
\usepackage{booktabs}         

\usepackage{soul}

\usepackage{setspace}
\graphicspath{{figures/}}

\usepackage{amsthm}
\usepackage{amsmath}
\usepackage{amsfonts}
\usepackage{cancel}

\usepackage{mathtools}
\DeclarePairedDelimiter{\nint}\lfloor\rceil

\newtheorem{theorem}{Theorem}
\newtheorem{definition}{Definition}

\usepackage{subcaption}

\usepackage{cleveref}
\crefname{equation}{Eq}{Eqs} 
\crefrangelabelformat{equation}{(#3#1#4--#5#2#6)}

\usepackage{float}

\usepackage{tikz}
\usetikzlibrary{backgrounds, calc, chains, fit, matrix, positioning, shadows, shapes, intersections, circuits.ee}
\tikzset{input/.style={}}
\tikzset{output/.style={}}
\tikzset{op/.style={circle, draw, fill=black!10, minimum size=2.5ex, inner sep=0ex}}
\tikzset{filter/.style={rectangle, draw, thick, fill=black!10, minimum size=3.5ex, inner sep=1ex}}
\tikzset{nn/.style={trapezium, trapezium angle=80, draw, thick, fill=black!10, inner sep=1ex}}
\tikzset{branch/.style={circle, draw, thick, fill=black, minimum size=.5ex, inner sep=0ex}}
\tikzset{tensor/.style={rectangle, draw, fill=white, minimum size=2em, double copy shadow={shadow xshift=.5ex,shadow yshift=-.5ex}}}
\tikzset{rounded/.style={rounded rectangle, draw, thick, fill=black!10, minimum size=3.5ex, inner xsep=1ex}}
\tikzset{image/.style={rectangle, draw, fill=white, minimum size=2em}}
\tikzset{>=direction ee}
\tikzset{/tikz/thin/.style={line width=.9pt}}
\tikzset{/tikz/thick/.style={line width=1.4pt}}
\tikzset{every path/.style={thin}}

\usepackage{pgfplots}
\pgfplotsset{compat=1.14}
\pgfplotsset{every axis/.append style={enlargelimits={abs=3pt},grid,axis lines=left}}
\pgfplotsset{every axis plot/.append style={thick,mark size=1.5pt,line join=bevel,mark options={solid}}}
\pgfplotsset{label style={font=\small}}
\pgfplotsset{tick label style={font=\footnotesize}}
\pgfplotsset{grid style={color=black!10}}
\pgfplotsset{legend style={draw=none,opacity=.85,font=\footnotesize,cells={anchor=west,opacity=1}}}
\pgfplotsset{every non boxed x axis/.style={xtick align=center,shorten <=-.5\pgflinewidth}}
\pgfplotsset{every non boxed y axis/.style={ytick align=center,shorten <=-.5\pgflinewidth}}
\pgfplotsset{every non boxed z axis/.style={ztick align=center,shorten <=-.5\pgflinewidth}}
\pgfplotsset{/pgf/number format/1000 sep={\,}}

\ifCLASSINFOpdf
\else
\fi
\begin{document}
%
\title{Neural Distributed Compressor \\
Discovers Binning}
%
%
%

\author{Ezgi~Ozyilkan,~\IEEEmembership{Graduate Student Member,~IEEE,}
        Johannes~Ballé,~\IEEEmembership{Member,~IEEE},~and~Elza~Erkip,~\IEEEmembership{Fellow,~IEEE}
\thanks{This work was presented in part at the IEEE International Symposium on Information Theory (ISIT), June 2023. This work was supported in part by NYU Wireless and Google Research Collabs Program. \emph{(Corresponding author: Ezgi Ozyilkan.)}}
\thanks{E. Ozyilkan and E. Erkip are with the Department
of Electrical and Computer Engineering, New York University, Brooklyn, NY, 11201 USA (e-mail: ezgi.ozyilkan@nyu.edu, elza@nyu.edu) .}
\thanks{J. Ballé is with Google Research, New York, NY 10011, USA (email: jballe@google.com).}
}

\doublespacing 
\onecolumn 

\maketitle

\begin{abstract}
  We consider lossy compression of an information source when the decoder has lossless access to a correlated one. This setup, also known as the \emph{Wyner--Ziv} problem, is a special case of distributed source coding. To this day, practical approaches for the Wyner--Ziv problem have neither been fully developed nor heavily investigated. We propose a data-driven method based on machine learning that leverages the universal function approximation capability of artificial neural networks. We find that our neural network-based compression scheme, based on variational vector quantization, recovers some principles of the optimum theoretical solution of the Wyner--Ziv setup, such as binning in the source space as well as optimal combination of the quantization index and side information, for exemplary sources. These behaviors emerge although no structure exploiting knowledge of the source distributions was imposed. Binning is a widely used tool in information theoretic proofs and methods, and to our knowledge, this is the first time it has been explicitly observed to emerge from data-driven learning.
\end{abstract}


%
\IEEEpeerreviewmaketitle

\section{Introduction}
\label{sec:intro}

\IEEEPARstart{C}{onsider} a distributed sensor network consisting of individual cameras that independently capture images at different locations across the same city. Suppose that each sensor node compresses and transmits its highly correlated image to a joint central processing unit that reproduces a unified visual map of the city, by fusing the information collected by all of the nodes. If the sensors could directly communicate with each other in a cooperative manner, they could avoid some degree of redundancy and compress more efficiently. However, direct communication between nodes is often infeasible.

Given that, what is the best strategy to exploit the correlation between sensor data? Slepian and Wolf~\cite{Slepian:IT:73} (SW) proved a remarkable and well-known information theoretic result that the distributed compression is asymptotically as efficient as the joint one, if the joint distribution statistics are known and compression is lossless. Their proof invokes \emph{random binning} arguments and is non-constructive. Establishing a practical framework building onto these concepts is a challenging open problem to this day.

Here, we investigate the setup characterized by Wyner and Ziv~\cite{Wyner:IT:76} (WZ), which is both more general than SW as it encompasses lossy compression, and a simpler special case, as it assumes the decoder has access to a correlated source, the \emph{side information}, directly. For WZ coding, there has been several prior work considering specific abstract source models. Zamir \emph{et al.}~\cite{zamir_ITW, zamir_TIT} outlined asymptotically optimal constructive mechanisms using nested linear and lattice codes for binary and Gaussian sources, respectively. Since then, the constructive and non-asymptotic research effort has been spearheaded by distributed source coding using syndromes (DISCUS)~\cite{DISCUS}, which formulated the WZ setup as a dual quantizer-channel coding problem. In a nutshell, the source is first quantized in a suitable manner according to its marginal density. Next, the quantization codebook space is partitioned into \emph{cosets}, according to the virtual channel arising between the side information and the quantized source. Instead of sending the quantization index to the decoder, the encoder only sends the index of the coset containing the quantized codeword, which results in further rate reduction. The decoder can then disambiguate the coset index with the help of the side information, and recovers the quantization index. Finally, it estimates the source using the deduced quantization index and the side information according to the distortion criterion. Note that such a systematic partitioning of the quantized source space with cosets effectively mimics the random binning procedure in the proofs of the SW and also of the WZ theorems that consider the asymptotic regime. The complex interaction between the quantization, channel coding and estimation parts was also highlighted in competitive practical code design frameworks proposed in~\cite{swc_nq, tcq_ldpc}. These methods achieve performances close to the theoretical bound, but are only explored with specific source distributions in mind, such as correlated Gaussians. WZ coding has also been investigated in the context of video compression~\cite{Girod}, where it has been shown that exploiting interframe dependence of a video sequence enables superior performance compared to conventional coding.

\begin{figure*}[t]
\centering
   \includegraphics[width=.6\linewidth]{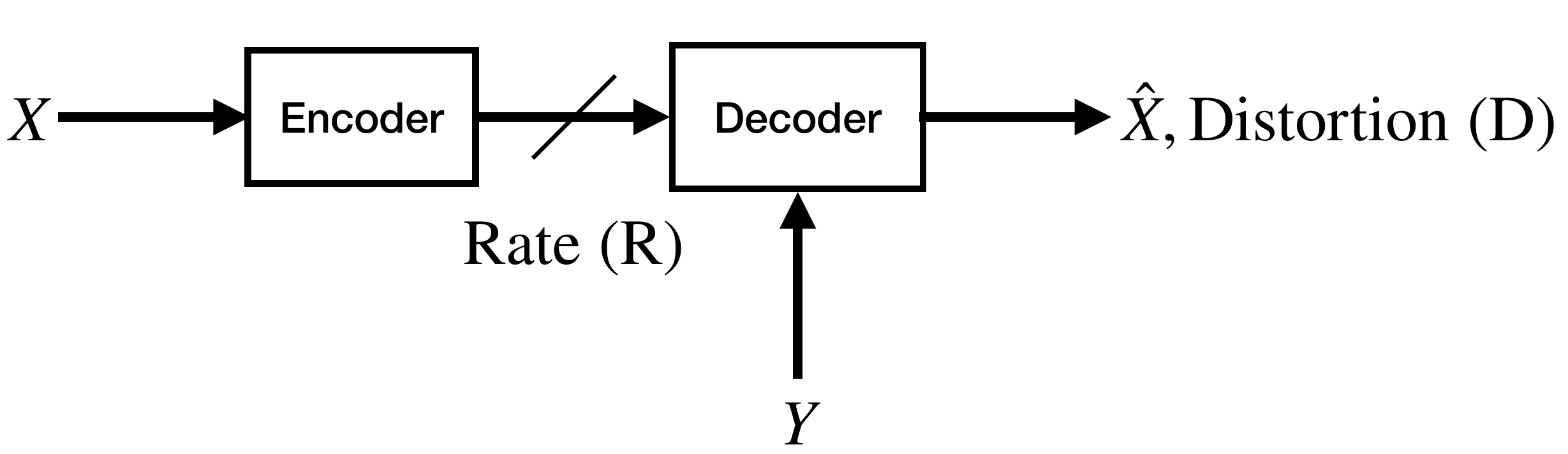}\caption{One-shot lossy source coding with decoder-only side information.}
  \label{fig:sys_initial} 
\end{figure*} 

In this work, we will leverage the universal function approximation capability of artificial neural networks (ANNs)~\cite{Leshno1993, hornik_et_al}  and machine learning techniques to find constructive solutions for the WZ problem in the non-asymptotic blocklength, particularly one-shot, regime (see Fig.~\ref{fig:sys_initial}). This is motivated by practice-oriented compression applications that require little delay and complexity, such as low-latency image and video transmission in real-time. We will also assume that the joint distribution of the source and side information is not known \emph{a priori}. As will be seen in detail, there are several immediate challenges when establishing a constructive ANN-based WZ compression scheme, which, ideally, should be able to exhibit adaptive binning mechanisms depending on the joint distribution of $X$ and $Y$. Although the popular class of neural methods based on stochastically-trained ANN-based compressors~\cite{balle_journal} seems to be good candidate for the source coding with side information setup we consider, as will be demonstrated, it fails to recover many-to-one mappings exploiting the side information and as a consequence, is unable to learn any proper binning scheme. We hypothesize that this is due to the ANNs having difficulty learning high frequency and discontinuous functions \cite{tancik2020fourier, ziyin2020neural}, a phenomenon referred to as \emph{spectral bias}~\cite{rahaman2019spectral, basri2020frequency} in the learning literature. We argue that this learning bias towards low frequency features prohibits this popular ANN-based compression framework from exploiting the side information most efficiently, and thus leads to suboptimal operational R-D performances even considering a simple use case of side information, which we examine closely in Section~\ref{sec:ntc_intro}. 

\subsection{Contributions}

As an alternative to the prevalent class of ANN-based compressors, we take a step back and propose a more generic learning-based algorithm based on vector quantization (VQ), for the source coding with side information setup. In particular, we neither impose any particular structure onto the model nor assume any prior knowledge about the source distributions. Specifically, unlike the popular set of ANN-based compressors, the proposed approach does \emph{not} operate in an ordered transform space on real line, but instead directly in an unordered \emph{categorical} one. Our main contributions can be summarized as follows.

\begin{itemize}
   \item We propose the first unstructured entropy-constrained VQ (ECVQ) that makes use of side information. We also provide post-hoc interpretations for the models by visualizing the behavior of the learned compressors.
   \item To the best of our knowledge, we are the first to demonstrate (with the help of illustrative example sources, various correlation patterns and visualizations) that a learned compressor is capable of exhibiting different types of interpretable binning mechanisms. 
  \item We present the empirical results of the proposed approach for some well-studied source distributions whose rate--distortion (R-D) bounds are known. We demonstrate that the learned decoder optimally combines the quantization index and side information for the quadratic-Gaussian case. We also show that our learned coder optimally compresses the Laplacian source with sign function as side information.
  \item The learned schemes we propose in this paper are simpler and more robust learning-based solutions compared to the ones in our previous work~\cite{ozyilkan2023learned}. Similarly to~\cite{ozyilkan2023learned}, we explain how each of our formulations corresponds to an operational scheme.
\end{itemize}

We motivate our design choice by presenting the popular class of stochastically-trained ANN-based compressors and analyze why these neural methods fail to incorporate side information in Section~\ref{sec:ntc_intro}. We introduce some basic concepts and definitions of relevance, establish our system setup, and summarize previous compressor designs relevant to our work in Section~\ref{sec:related_work}. Next, we introduce our novel neural framework, based on variational ECVQ, for the WZ problem, and also justify how each model is interpretable as an operational scheme in Section~\ref{sec:method}. We elaborate on our experimental setup in Section~\ref{sec:experimental_setup}. Finally, we discuss empirical results and connections to related work, and conclude the paper in Section~\ref{sec:discussion}.

\section{Popular ANN-Based Compressor Fails to Exploit Side Information}
\label{sec:ntc_intro}

Most popular previous work on end-to-end learned lossy compression literature can be collected under the banner of nonlinear transform coding (NTC) \cite{balle_journal}. Inheriting a divide-and-conquer paradigm from its conceptual ancestor, that is transform coding (TC)~\cite{goyal}, NTC greatly simplifies the joint optimization of rate and distortion by mapping the source into a \emph{latent space} using learnable nonlinear decorrelating transforms, and then individually quantizing and coding each of the dimensions in the latent space. Note that differently from traditional VQs, where the codebook vectors are constructed explicitly, in NTC case, the codebook is implicitly governed by the learned transforms (see the discussion in Section 3 in~\cite{balle_journal}). NTC-based compression schemes can easily adapt to any arbitrary source distributions, simply by replacing the training data, as well as to any differentiable distortion measures, through end-to-end stochastic optimization methods. Recently, NTC-based models have superseded the best linear transform codecs for images (such as JPEG \cite{JPEG}), under both traditional and perceptual quality metrics~\cite{balle2018variational}. Furthermore, it has been also shown that NTC-based schemes can optimally compress sources that even exhibit a low-dimensional manifold structure in a high-dimensional ambient space, such as a specific random process named the \emph{sawbridge}~\cite{wagner2020neural}.

In the general case of point-to-point compression (no side information) setup, the neural compression models based on this data-driven NTC framework optimize an R-D objective of the following form:
\begin{equation}
    -\log_2 q_{\boldsymbol{\psi}}(\nint{(f_{\boldsymbol{\theta}}(\mathbf{x})}) + \lambda \cdot d(\mathbf{x}, g_{\boldsymbol{\phi}}(\nint{(f_{\boldsymbol{\theta}}(\mathbf{x})}))\footnote{Throughout, all logarithms are base two.}.
    \label{eq:classic_rd}
\end{equation}Here, $d(\cdot, \cdot)$ is a distortion measure that quantifies the discrepancy between inputs and reconstructions; $f_{\boldsymbol{\theta}}$ and $ g_{\boldsymbol{\phi}}$ are learned encoder (\emph{analysis}), $f_{\boldsymbol{\theta}} : \mathbb{R}^{n_s} \mapsto \mathbb{R}^{n_l}$, and decoder (\emph{synthesis}) $g_{\boldsymbol{\phi}} : \mathbb{R}^{n_l} \mapsto \mathbb{R}^{n_s}$ functions, with parameters $\boldsymbol{\theta}$ and $\boldsymbol{\phi}$, respectively. Also, $\nint{\cdot}$ denotes uniform scalar quantization (rounding to integers), and $q_{\boldsymbol{\psi}}$ corresponds to the learned probability distribution model, with parameters $\boldsymbol{\psi}$, over the quantized latents. Since the quantization operation is not differentiable, Agustsson and Theis~\cite{agustsson2020universally} propose to replace it with a differentiable proxy over the course of training, which begins with dithered quantization and ends with the hard quantizer, coinciding with what is actually used at test time. During training, a $n_s$-dimensional vector is fed into the analysis transform, which outputs a $n_l$-dimensional latent vector. The aforementioned quantization proxy is applied to this latent vector, which is subsequently fed into the synthesis transform to obtain a $n_s$-dimensional reconstruction vector\footnote{Similarly to its TC counterpart, NTC-based quantization scheme is defined indirectly through the analysis and synthesis transforms, where the former dictates the quantization boundaries and the latter determines the effective codebook vectors.}.

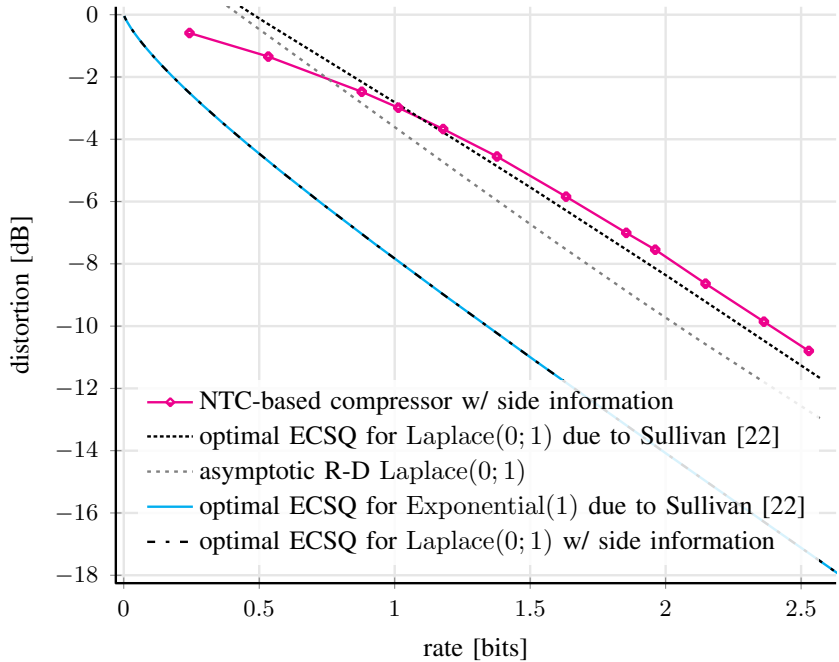
\begin{figure*}[t]
    \centering
    \begin{tikzpicture}[trim axis right]
    \begin{axis}[
      height=.33\textheight,
      width=.58\linewidth,
      scale only axis,
      xlabel={rate [bits]},
      ylabel={distortion [dB]},
      xmin=0.,
      xmax=2.6,
      ymin=-18.,
      ymax=0.,
      legend pos=south west,
      legend style={font=\small},
      ]
      \addplot[color=magenta, mark=halfsquare*] table{
      0.2426915 -0.5877767130732536
      0.53348315 -1.3451550900936127
      0.8783622 -2.474716752767563
      1.0134085 -2.9860273003578186
      1.1787941 -3.6717602610588074
      1.3775315 -4.553080797195435
      1.6324406 -5.842599272727966
      1.8546665 -7.005162239074707
      1.9615476 -7.54610538482666
       2.1473196 -8.637049198150635
       2.3627374 -9.857673048973083
       2.528376 -10.79639196395874
      };
      \addplot[color=black, densely dotted] table {
      2.56751 -11.655000043360188
      2.00991 -8.417500043360187
      1.50291 -5.556480043360187
      1.26059 -4.222570043360188
      1.05281 -3.095010043360188
      0.76978 -1.575450043360188
      0.51619 -0.21139004336018807
      0.250851 1.2708499566398122
      0.106253 2.163196956639812
      0.0551177 2.521273956639812
      0.00238992 2.977434956639812
      };
       \addplot[color=gray, dotted] table {
       2.56751 -12.947800043360187
       2.13267 -10.499400043360188
       1.8955 -9.117600043360188
       1.59269 -7.297800043360187
       1.26059 -5.243130043360187
       1.05281 -3.940480043360188
       0.820686 -2.482050043360188
       0.632445 -1.300720043360188
       0.449202 -0.14804004336018783
       0.337394 0.563029956639812
       0.270387 0.9960199566398122
      };
      \addplot[color=cyan] table {
        2.7742027930105997 -18.785274783850056
        1.9968246064495052 -14.059477010209605
        1.515411082139869 -11.093318831718413
        1.1770338248299443 -8.968760154307365
        0.9248340090286805 -7.346774483304259
        0.7309059225521122 -6.063032646142112
        0.5791536438400673 -5.024449745589495
        0.4592349040357366 -4.172502040049563
        0.3639795120626874 -3.467498912574212
        0.28814109566804225 -2.880928884872328
        0.22773403540136747 -2.391378957669089
        0.1796528896370207 -1.9822023028459168
        0.1414383079876411 -1.6401087653143276
        0.11112348127838338 -1.3542720666872214
        0.08712729304083434 -1.1157403055998345
        0.06817605103829918 -0.9170315326904719
        0.053243736271954153 -0.7518460510137284
        0.041504978885045715 -0.6148545278545412
        0.03229728782946271 -0.5015367192135871
        0.02509034126919169 -0.40805491911261704
        0.010533546063756546 -0.19889235770449876
        0.005856352988695899 -0.12130092798325709
        0.0017806358848948879 -0.04364218379077078
      };
      \addplot[color=black, loosely dashdotted] table {
        2.7742027930105997 -18.785274783850056
        1.9968246064495052 -14.059477010209605
        1.515411082139869 -11.093318831718413
        1.1770338248299443 -8.968760154307365
        0.9248340090286805 -7.346774483304259
        0.7309059225521122 -6.063032646142112
        0.5791536438400673 -5.024449745589495
        0.4592349040357366 -4.172502040049563
        0.3639795120626874 -3.467498912574212
        0.28814109566804225 -2.880928884872328
        0.22773403540136747 -2.391378957669089
        0.1796528896370207 -1.9822023028459168
        0.1414383079876411 -1.6401087653143276
        0.11112348127838338 -1.3542720666872214
        0.08712729304083434 -1.1157403055998345
        0.06817605103829918 -0.9170315326904719
        0.053243736271954153 -0.7518460510137284
        0.041504978885045715 -0.6148545278545412
        0.03229728782946271 -0.5015367192135871
        0.02509034126919169 -0.40805491911261704
        0.010533546063756546 -0.19889235770449876
        0.005856352988695899 -0.12130092798325709
        0.0017806358848948879 -0.04364218379077078
      };
      \legend{NTC-based compressor w/ side information, optimal ECSQ for $\mathrm{Laplace}(0;1)$ due to Sullivan~\cite{sullivan}, asymptotic R-D $\mathrm{Laplace}(0;1)$, optimal ECSQ for $\mathrm{Exponential}(1)$ due to Sullivan~\cite{sullivan}, optimal ECSQ for $\mathrm{Laplace}(0;1)$ w/ side information};
    \end{axis}
    \end{tikzpicture}
     \caption{Rate--distortion performances (R-D) obtained with Nonlinear Transform Coding (NTC)~\cite{balle_journal}, which is adapted to incorporate the available side information, that uses a variant of Eq.~\eqref{eq:classic_rd} as the objective function. Here, we consider a simple one-shot source coding with side information setting, where $X \sim \mathrm{Laplace}(0;1)$  and $Y= \mathrm{sgn}(X)$, i.e., the sign function of the input realization. We also plot the operational R-D function of the optimal entropy-constrained scalar quantizers (ECSQs), due to Sullivan~\cite{sullivan}, for the Laplacian and exponential sources. Note that the optimal ECSQ for the Laplacian source having sign information available at the decoder coincides with the optimal ECSQ for the exponential source.}
     \label{fig:ntc_initial}
\end{figure*}

One can easily adapt the objective function in Eq.~\eqref{eq:classic_rd} in order to fuse the side information by replacing the decoding function with $g_{\boldsymbol{\phi}}(\nint{(f_{\boldsymbol{\theta}}(\mathbf{x})}), \mathbf{y})$, which now instead receives a concatenated vector of both inputs. A similar line of reasoning was followed by the image compression works in~\cite{NDIC, NDIC_CAM}, where an NTC-based scheme was used to exploit the decoder-only side information in the form of a correlated image.

As a simple test case for this NTC-based scheme with side information (see Fig.~\ref{fig:sys_initial}), we consider the following one-shot source coding with side information setup: let $X \sim \mathrm{Laplace}(0;1)$  and $Y= \mathrm{sgn}(X)$, i.e., the sign function of the input realization, and let the distortion metric $d(\cdot, \cdot)$ be MSE. Note that since $Y$ is a deterministic function of $X$, the optimal encoder would simply correspond to the one that compresses $|X| \sim \mathrm{Exponential}(1)$. Because the source is of dimension one in this one-shot case, i.e., $n_s = 1$, we set the latent dimension to be also one, i.e., $n_l = 1$.

By sweeping across different values of $\lambda$ (where we use Eq.~\eqref{eq:classic_rd} with the side information and a modified version of $g_{\boldsymbol{\phi}}$), we obtain different points in the achievable R-D region and plot them in Fig.~\ref{fig:ntc_initial}. As we discuss in Section~\ref{subsec:system_setup}, we can use the entropy-constrained scalar quantization (ECSQ)~\cite{sullivan} framework to benchmark performance of NTC-based compressors. Although the ECSQ bounds assume a specific joint distribution of $p(x,y)$ in their formulation, the NTC framework does \emph{not} make such an assumption and learns the analysis and synthesis transforms through training. As argued above, under this correlation structure, the optimum ECSQ with side information performance corresponds to that of the ECSQ with exponential distribution. The points plotted for the asymptotic R-D bound were computed using the Blahut-Arimoto algorithm~\cite{blahut, arimoto} using a proper discretization of the source.  As seen, in low rates, the NTC-based framework achieves a slightly better performance than the optimal ECSQ that compresses $X$ without any side information. This observation is consistent with the one presented in~\cite{NDIC, NDIC_CAM}, where Mital \emph{et al.} considered a pair of stereo images as correlated sources in the low rate regime. At higher rates, however, the NTC-based compressor performs no better than the point-to-point ECSQ. Inspecting the transform functions learned by the NTC-based models optimized for each $\lambda$ values, we find no evidence of binning, or more generally many-to-one mappings, occurring in the source space. We also find that the learned decoder does not make use of the side information. We attribute the suboptimal performance of this NTC-based model to its incapability of learning high frequency functions that are necessary to recover a variant of binning, which is arguably caused by the ANN-based compressor's \emph{spectral bias}. Specifically, it has been demonstrated in the learning literature that ANNs prioritize learning low frequency modes, which renders them biased towards learning smooth functions instead~\cite{rahaman2019spectral}. The learning difficulty in this case may be linked to the factor that in the NTC paradigm, the latent elements on real line are rounded to the nearest integer values. This may create a topological challenge for the model to recover many-to-one mappings and high frequency functions in such an \emph{ordered} latent space, in order to efficiently make use of the available side information. This observation is analogous with the one in~\cite{bhadane2022neural}, where Bhadane \emph{et al.} observe that the stochastically-trained NTC-based compressor fails to learn sufficiently steep functions, which leads to distortion at the discontinuities. As a consequence, they demonstrate that this popular NTC-based model is unable to optimally compress some complex sources exhibiting circular topology, such as a particular random process named the \emph{ramp}.

\section{Preliminaries, Setup and Prior Work}
\label{sec:related_work}

In Section~\ref{sec:rd_with_side_info}, we briefly summarize relevant theoretical results on the lossy source coding side. In Section~\ref{sec:optimal_operational_rd_function}, we provide some definitions related to the operational R-D performance achieved by practical compressors, upon which we will later build our formulation. We briefly lay out our system setup in Section~\ref{subsec:system_setup}. We finish by mentioning some related compressor designs in the literature in Section~\ref{sec:relevant_quantizer_designs}.

\subsection{Asymptotic Rate--Distortion Functions}
\label{sec:rd_with_side_info}

Let $X^{n}$ be a sequence, drawn i.i.d. $\sim p(x)$ for $i=1, ..., n$; and let $d(x, \hat{x})$ be a single letter distortion measure. By the well-known lossy source coding theorem~\cite{network_info_theo}, the point-to-point R-D function, which assumes asymptotically large blocklengths $n$, is given as:
\begin{equation}
      R(D) \, = \, \min_{p(\hat{x} \vert x) \; : \;  \mathbb{E}(d(x,\hat{x})) \leq D} I(X;\hat{X}). \label{eq:rd_p2p}
\end{equation} 

\begin{figure*}[t]
\centering
   \includegraphics[width=.55\linewidth]{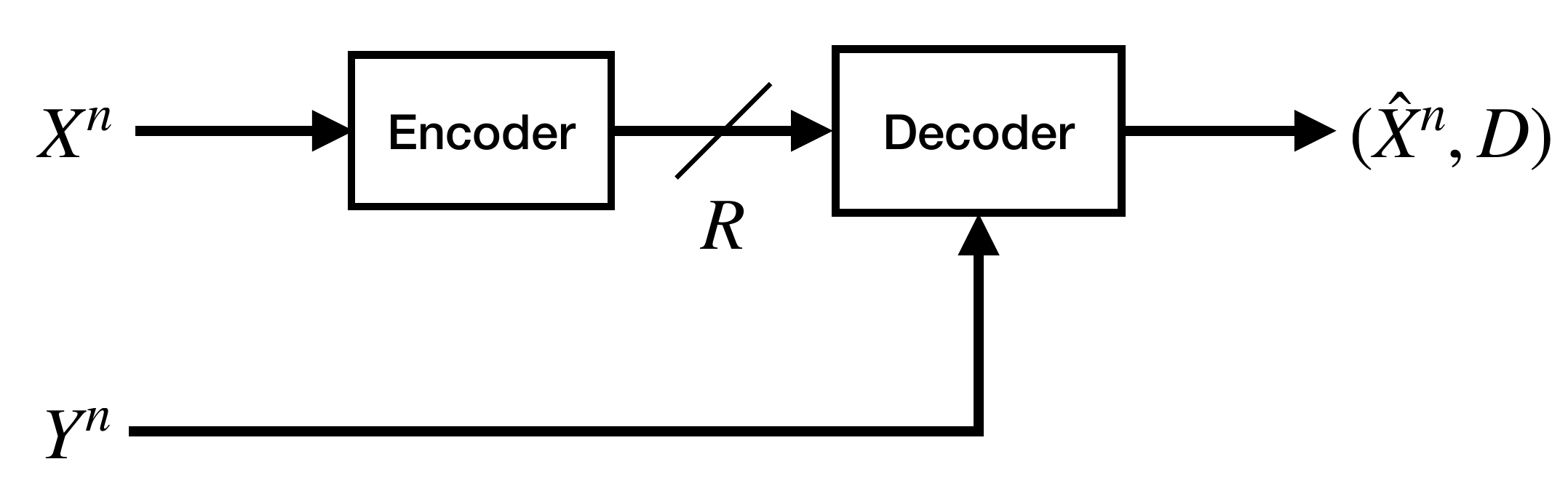}
\caption{Lossy source coding with decoder-only side information in the asymptotic blocklength regime.}
  \label{fig:sys_side_info} 
\end{figure*}

Now, let $(X^{n},Y^{n})$ be correlated sequences, drawn i.i.d. $\sim p(x,y)$ for $i=1,...,n$. As in Fig.~\ref{fig:sys_side_info}, the source $X^{n}$ is compressed and sent over a noiseless link, with a 
rate of $R$ bits/source sample, to a decoder that has side information $Y^{n}$ whose goal is to reproduce the source within average distortion $D$. Note that the side information is available in a \emph{noncausal} manner only at the decoder, which is also known as the WZ setup~\cite{Wyner:IT:76}. The R-D function for this case, which is achieved in the limit of $n \rightarrow \infty$, is given by the following theorem. For the complete proof, we refer readers to the original paper~\cite{Wyner:IT:76} and to~\cite{network_info_theo}.

\begin{theorem} \label{theo:WZ}(Wyner--Ziv Theorem [1976]) Let $(X,Y)$ be correlated sources, drawn i.i.d. $\sim p(x,y)$, and let $d(x, \hat{x})$ be a single-letter distortion measure. The R-D function for $X$ with side information $Y$
available (noncausally) at the decoder is as follows:
\begin{equation} \label{eq:WZ}
    R_{\text{WZ}}(D) \, = \, \min (I(X;U) - I(Y;U)),
\end{equation}
where the minimization operation is over
all conditional probability distribution functions $p(u \vert  x)$,
and all functions $g(u,y)$ such that  $ \mathbb{E}_{p(x,y)p(u\vert x)} d(x, g(u,y) ) \le D$. \end{theorem}The achievability part of the WZ theorem invokes the covering lemma, resulting in the compression rate of $I(X;U)$, followed by a random binning argument based on joint typicality, which yields the rate discount of $I(Y;U)$ in Eq.~\eqref{eq:WZ}~\cite{elements_of_information_theory}. This achievability, which is shown to be tight, assumes a Markov chain constraint $U-X-Y$.

As in classical R-D theory, Theorem~\ref{theo:WZ} assumes asymptotically large blocklengths. However, practical compression schemes need to operate with a finite number of source samples. One solution to overcome this is VQ followed by variable length binary coding the quantization indices, i.e., \emph{entropy encoding}, through which the average transmission rate is reduced to the entropy rate of the index sequence. As such, the induced rate is no longer the logarithm of the codebook size but rather the average length of the binary codewords, where the average is over the distribution of the source samples~\cite{quantization}. This scheme is known as \emph{ECVQ} in the compression literature~\cite{ECVQ}. To achieve the desired coding rate, most practical compression schemes have a \emph{variable rate} nature, hence involving some added complexity in the system design, which we look at next.

\subsection{Operational Rate--Distortion Functions} \label{sec:optimal_operational_rd_function}

In order to provide operational R-D bounds to our proposed neural compressors, we first define a pair of point-to-point one-shot encoder and decoder, and their entropy and distortion, in a similar exposition style as the ones in~\cite{wagner2020neural, bhadane2022neural}. As we explain in Section~\ref{subsec:system_setup}, our proposed neural compressors operate in the one-shot regime.

\begin{definition}
Consider a source $X \sim p(x)$, $x \in \mathcal{X}$, and a distortion function $d(x, \hat{x})$, $\hat{x} \in \hat{\mathcal{X}}$, where $\mathcal{X}$ and $\hat{\mathcal{X}}$ are source and reconstruction alphabets, respectively.
A one-shot lossy encoder is a mapping $f_{p} : \mathcal{X} \to \mathbb{N}$, and its associated decoder is a mapping $g_{p}: \mathbb{N} \to \hat{\mathcal{X}}$. The \textup{entropy} and \textup{distortion}, induced by such a pair of \textup{encoder} and \textup{decoder}, are given by
\begin{equation}
    H(f_{p}(X)) = - \sum_{i \in \mathbb{N}}\mathrm{Pr}(f_{p}(X)=i)\log (\mathrm{Pr}(f_{p}(X)=i)),
    \label{eq:entropy_formal}
\end{equation}
\begin{equation}
    D(f_{p}(X), g_{p}(X)) = \mathbb{E}_{p(x)} \left[d(X, g_{p}(f_{p}(X))\right],
    \label{eq:distortion_formal}
\end{equation}
respectively.
\end{definition}
\noindent In our case, $d(\cdot, \cdot)$ in Eq.~\eqref{eq:distortion_formal} can be any differentiable quantity since the neural compressors can be optimized for any differentiable distortion measure via stochastic optimization techniques. In the traditional quantization literature, the most common distortion measure is the squared error, i.e., $d(x,\hat{x})=(x-\hat{x})^2$. In this case, the optimal decoding function is simply the conditional mean, and therefore, the Eq.~\eqref{eq:distortion_formal} can be replaced, without loss of generality, directly with:
\begin{equation}
    D(f_{p}(X)) = \mathbb{E}_{p(x)} \left[ \parallel X - \mathbb{E}[X \, \vert \,  f_{p}(X)] \parallel^{2}\right].
\end{equation}

We remark that what (point-to-point) ANN-based compressors (e.g., NTC) and ECSQs optimize, in fact, coincides with \emph{Shannon entropy} of the quantized encoder output (see Eqs.~\eqref{eq:classic_rd} and~\eqref{eq:entropy_formal}), which is known to be a lower bound to the expected codeword length under optimal, one-shot prefix-free encoding~\cite{elements_of_information_theory}. In addition, it is also known that this lower bound is achievable if several encoded indices are compressed losslessly at once, while encoding/quantization (which often is the more complex operation) is kept as one-shot~\cite{quantization}. We can now introduce the optimal operational one-shot R-D performance via the \emph{entropy--distortion function}.
\begin{definition}~\cite[Definition 3]{wagner2020neural}~\cite[Definition 2]{bhadane2022neural} \label{def:E_D}
The entropy--distortion function of source $X \sim p(x)$ with distortion function $d(\cdot, \cdot)$ is
\begin{equation*}
    E(\Delta) = \inf_{f_{p},g_{p}} H(f_{p}(X)),
\end{equation*}
where the infimum is over all encoders $f_{p}$ and decoders $g_{p}$ such that $D(f_{p}(X), g_{p}(X)) \leq \Delta$.
\end{definition} For a correlated source $(X,Y)$ in a space $\mathcal{X} \times \mathcal{Y}$, we can similarly define a one-shot encoder and decoder that has access to side information, as well as their induced entropy and distortion. Since we will consider one-shot lossy compression of continuous sources, $\mathcal{X}, \hat{\mathcal{X}}$ and $\mathcal{Y}$, will be all $\mathbb{R}^{1}$ in this work.

\begin{definition} \label{def:e_d_with_side_info}
Consider a source $(X, Y) \sim p(x,y)$, $x \in \mathcal{X}$ and $y \in \mathcal{Y}$, a distortion function $d(x, \hat{x})$, $\hat{x} \in \hat{\mathcal{X}}$, where $\mathcal{X}, \mathcal{Y}$, and $\hat{\mathcal{X}}$ are source, side information, and reconstruction alphabets, respectively. A pair of one-shot lossy encoder and decoder, where the decoder has access to side information $Y$, consists of mappings $f_{s} : \mathcal{X} \to \mathbb{N}$ and $g_{s}: \mathbb{N} \times \mathcal{Y} \to \hat{\mathcal{X}}$, respectively. The \textup{entropy} and \textup{distortion}, induced by such a pair of \textup{encoder} and \textup{decoder}, are given by
\begin{equation}
    H(f_{s}(X) \; \vert \;  Y) = - \mathbb{E}_{p(y)} \Big[\sum_{i \in \mathbb{N}}\mathrm{Pr}(f_{s}(X)=i \; \vert \;  Y=y)\log (\mathrm{Pr}(f_{s}(X)=i \; \vert \;  Y=y)) \Big],
    \label{eq:entropy_side_info_formal}
\end{equation}
\begin{equation}
    D_{\mathrm{SI}}\big(f_{s}(X), g_{s}(f_{s}(X),Y)\big) = \mathbb{E}_{p(x,y)} \left[d\left(X, g_{s}(f_{s}(X), Y\right)\right],
    \label{eq:distortion_side_info_formal}
\end{equation}  
respectively.
\end{definition} Similar to the point-to-point case, we can now introduce the optimal operational one-shot R-D performance with decoder-side information via the \emph{entropy--distortion function with decoder-only side information}. 
\begin{definition} \label{def:E_D_side_info}
The entropy--distortion function with decoder-only side information for a correlated source $(X,Y)$ is
\begin{equation*} \label{def:entropy_opt_side_info}
    E_{\mathrm{SI}}(\Delta) = \inf_{f_{s},g_{s}} H(f_{s}(X) \; \vert \; Y),
\end{equation*}
where the infimum is over all encoders $f_{s}$ and decoders $g_{s}$ such that $ D_{\mathrm{SI}}\big(f_{s}(X), g_{s}(f_{s}(X),Y)\big) \leq \Delta$.
\end{definition}

While Eq.~\eqref{eq:entropy_side_info_formal} still provides a lower bound to the expected codeword length of any code with decoder only side information, unlike the point-to-point case, there are no known practical entropy coders achieving this conditional entropy for an arbitrary source distribution of $p(x,y)$ (see Section~\ref{subsec:operational_schemes} for the relevant discussion). Note that Eq.~\eqref{eq:entropy_side_info_formal} corresponds to the SW bound~\cite{Slepian:IT:73}, and many works prior works assume the availability of either an ideal or a practical SW coder (SWC) in their design (see Section~\ref{sec:relevant_quantizer_designs} for the relevant discussion).

\subsection{Our System Setup} \label{subsec:system_setup}
We propose to leverage the universal function approximation capability of ANNs~\cite{Leshno1993, hornik_et_al} and stochastic optimization techniques to find constructive solutions for the non-asymptotic regime for the WZ setup. More specifically, we consider the \emph{one-shot} case, i.e., compressing each source realization one at a time (see Fig.~\ref{fig:sys_initial}) followed by \emph{high-order} entropy coding over large blocks of the quantized source, similarly to popular stochastically-trained ANN-based compressors (e.g., NTC)~\cite{Balle2017, balle2018variational, balle_journal} and practical compressor designs in the literature (e.g., ECSQ~\cite{sullivan}). We do not assume \emph{a priori} knowledge of $p(x,y)$, and only assume samples from the distributions are available for training purposes.

In order to perform close to the entropy--distortion function that incorporates side information as in Definition~\ref{def:entropy_opt_side_info}, we will specifically provide two distinct solutions to the WZ problem, where we either handle the quantization and binning parts jointly and have a classic entropy coder or take a two-step approach by having a learned quantizer that is coupled with an ideal SWC, both of which we will explain in detail in Section~\ref{subsec:operational_schemes}.

We are not aware of any one-shot theoretical bounds with decoder side information in the literature that consider variable-length source coding. Kostina and Verdù~\cite{Kostina_2012} provide achievability results in the finite blocklength regime for the fixed-length compression setting. The most recent one-shot achievability result for the WZ setting in~\cite{Li_2021} also considers fixed-length compression and uses the \emph{Poisson matching lemma}. Hence, our performance benchmarks will include the entropy--distortion function with decoder-only side information, which we introduce in Definition~\ref{def:E_D_side_info}, or the (asymptotic) WZ R-D function, depending on the availability of each bound for the sources we consider.

\subsection{Relevant Lossy Compressor Designs in the Literature} \label{sec:relevant_quantizer_designs}

Designing both optimal or heuristic compressors with a focus on practical sources, such as image and speech, has been an active area of research for decades. In general, the entropy induced by a \emph{fixed rate} quantizer\footnote{A quantizer is said to have \emph{fixed rate} when all quantizer levels are assumed to have binary codewords of equal fixed-length~\cite{quantization}.} is solely dictated by the number of partitions it allows in the source space. The design of such fixed rate quantizers then amounts to finding the partitions and the associated reproductions of the source vectors, which should result in minimum distortion between the source itself and its quantized version. The well-known \emph{Lloyd--Max algorithm}\footnote{This well-known class of iterative algorithms was independently developed by S.P. Lloyd and J. Max, in 1957 and 1960, respectively. Lloyd's work was done in Bell Laboratories and, although widely being circulated, was not published until 1982~\cite{Lloyd}. Max's work was published in 1960~\cite{Max}.} results in such a fixed rate quantizer, providing efficient methods to obtain optimum partitions as well as their reproductions. Later, the fixed rate quantizer design is generalized to account for entropy coding, thus allowing a more flexible R-D trade-off, as in~\cite{Berger1, Berger2, Wood}, and yielding what is known as \emph{entropy-constrained} quantizer design. Building onto Definition~\ref{def:E_D} and typically having MSE as the distortion metric, they minimize an objective function in the form of the Lagrangian $E+\lambda D$.

The optimal ECSQ design has been studied specifically for (negative) exponential and Laplacian sources by Sullivan in~\cite{sullivan}, where he proposed noniterative algorithms for obtaining optimal schemes by exploiting the \emph{memoryless} property of the exponential distribution. In~\cite{Gyorgy}, the optimal ECSQ design is established for a uniform source distribution considering a wide class of difference distortion measures. Furthermore, an ECSQ design for Gaussian input sources was studied in~\cite{Marco}, but the performance has been only characterized for the low-resolution region where rate is small.

Early attempts to design quantizers specific to the WZ setting were based on high-dimensional nested lattice constructions, followed by either fixed rate~\cite{zamir_ITW} or variable rate~\cite{servetto2006lattice} coding. Later, the non-asymptotic research effort for the WZ setup has been rekindled with the seminal DISCUS framework~\cite{DISCUS}, where Pradhan and Ramchandran proposed a scalar quantization step followed by an SW binning scheme. They borrowed ideas from the channel coding literature (e.g., syndromes of a linear channel code) for coset constructions at the intermediate output of the quantizer. We highlight that the focus of DISCUS was \emph{not} to establish an optimum quantizer design, but rather to come up with a \emph{constructive} practical WZ compression scheme that has computationally efficient encoding and decoding algorithms, which are, however, constructed for specific source distributions in mind (such as correlated Gaussians). More recently, an ECSQ framework that incorporates decoder-only side information, again only suited for Gaussian sources, is proposed in~\cite{ecsq_w_side_info}. This quantizer design with decoder side information builds onto the approach followed by DISCUS, which consists of a scalar quantizer followed by either an ideal or a practical SWC. Proposing a scheme based on Lloyd type I algorithm~\cite{Lloyd}, Tu \emph{et al.} establishes an iterative algorithm to numerically find the optimum partitions. In the technical report available in~\cite{Dalton}, it is shown that one-dimensional nested scalar quantization followed by SW coding can achieve up to $1.53$ dB from the asymptotic WZ bound at high rates, for the quadratic-Gaussian case. Note that this coincides with the same performance gap~\cite{quantization} between that of an optimal point-to-point scalar quantization and asymptotic R-D bound with no side information (see Eq.~\eqref{eq:rd_p2p}). We discuss connections to related work on the machine learning side, such as~\cite{ozyilkan2023learned, NDSC}, in Section~\ref{subsec:discussion}.

\section{Neural upper bounds on Wyner--Ziv and Operational Schemes}
\label{sec:method}

In this section, we establish the training objectives for learning-based ECVQ schemes we propose for the WZ problem. As will be presented in Section~\ref{subsec:estimating_bounds}, our upper bounds include minimization on the optimal entropy--distortion function with side information (see Definition~\ref{def:entropy_opt_side_info}). We also introduce learnable encoder and decoder functions for these variational ECVQ schemes and accompanying parametrization of the neural models in Section~\ref{subsec:estimating_bounds}. Next, we explain how each system model is interpretable as an operational scheme in Section~\ref{subsec:operational_schemes}.

\subsection{Estimating Neural Upper Bounds on Wyner--Ziv} \label{subsec:estimating_bounds}
Our goal is to find constructive solutions for the WZ problem, in the one-shot compression regime. However, unlike many prior works (see Section~\ref{sec:relevant_quantizer_designs}), we will not rely on any knowledge about source distributions, but will only assume 
an ability to sample from the distributions.

Unlike our previous work~\cite{ozyilkan2023learned}, in our new formulation we directly use a deterministic encoder. Specifically, we set the encoder as $U \triangleq f_{s}(X) $ as in Definition~\ref{def:e_d_with_side_info}. We also have $U$ as discrete as in our previous work~\cite{ozyilkan2023learned}. Note that any choice of $f_{s}(\cdot)$, and hence $U$, provides an upper bound to the entropy--distortion function with side information (see Definition~\ref{def:entropy_opt_side_info}) as well as an upper bound to the WZ R-D function (see Theorem~\ref{theo:WZ}). Building onto Definition~\ref{def:entropy_opt_side_info}, for our objective functions, we choose one of two variational upper bounds:
\begin{align}
    H(U \; \vert \; Y)  & \leq H(U)  \leq \mathbb{E}_{p(x)} \Big[-\log{q_{\boldsymbol{\xi}} (u)} \Big]_{u=f_{s}(x)}, \label{eq:upper_bound_marg_2} \\
    H(U \; \vert \; Y) &\leq \mathbb{E}_{p(x,y)} \Big[-\log{q_{\boldsymbol{\zeta}} (u \vert  y)}\Big]_{u=f_{s}(x)}.  \label{eq:upper_bound_cond_2}
\end{align} Here, $q_{\boldsymbol{\xi}}(u)$ and $q_{\boldsymbol{\zeta}}(u \vert y)$ (with parameters $\boldsymbol \xi$ and $\boldsymbol \zeta$, respectively), are two different models of the distribution $p(u)$ and $p( u \vert y)$, respectively, which are generally not known in closed form. We defer the discussion of the operational meaning of these two variants, including classic entropy and SW coding for Eqs.~\eqref{eq:upper_bound_marg_2} and \eqref{eq:upper_bound_cond_2} respectively, to Section~\ref{subsec:operational_schemes}. The upper bounds in Eqs.~\eqref{eq:upper_bound_marg_2} and \eqref{eq:upper_bound_cond_2} follow from cross-entropy~\cite{kullback} being larger or equal to entropy~\cite[Theorem 5.4.3]{elements_of_information_theory}.

Next, we relax the constrained formulation of the WZ theorem to an unconstrained one using a Lagrange multiplier. Building on the upper bounds in Eqs.~\eqref{eq:upper_bound_marg_2} and \eqref{eq:upper_bound_cond_2}, this yields either a \emph{marginal} or a \emph{conditional} loss function:
\begin{align} \label{eq:proposed_loss_marginal}
     L_\mathrm{m}(\boldsymbol{\xi}, \boldsymbol{\phi}) &= \mathbb{E}_{p(x,y)} \Big[-\log q_{\boldsymbol{\xi}}(u) + \lambda  \cdot  d(x, g_{\boldsymbol{\phi}}(u, y))\Big]_{u=f_{m}(x)} \; , \\ \label{eq:proposed_loss_conditional}
     L_\mathrm{c}(\boldsymbol{\boldsymbol{\boldsymbol{\zeta}}}, \boldsymbol{\phi}) &= \mathbb{E}_{p(x,y)} \Big[-\log q_{\boldsymbol{\zeta}}(u \vert y) + \lambda \cdot d(x, g_{\boldsymbol{\phi}}(u, y))\Big]_{u=f_{c}(x)} \; ,
\end{align} where $\{\boldsymbol{\xi},
\boldsymbol{\phi},  \boldsymbol{\zeta}\}$ are optimization parameters. Here,
$f_{m}(\cdot)$ and $f_{c}(\cdot)$ correspond to the respective encoding functions for these two variants, and $g_{\boldsymbol{\phi}}(u,y)$ denotes the decoding function, represented by an ANN with parameters $\boldsymbol{\phi}$, which outputs the reconstruction $\hat{x}=g_{\boldsymbol{\phi}}(u, y)$. Before we explain how the learnable parameters $\{\boldsymbol{\xi},
\boldsymbol{\phi},  \boldsymbol{\zeta}\}$ are optimized, we discuss the parametrization of the models $q_{\boldsymbol{\xi}}(u)$ and $q_{\boldsymbol{\zeta}}(u \vert y)$, and introduce the encoding functions $f_{m}(\cdot)$ and $f_{c}(\cdot)$.

Similarly to~\cite{ozyilkan2023learned}, without loss of generality, we define all probabilistic models $q_{\boldsymbol{\xi}}(u)$ and $q_{\boldsymbol{\zeta}}(u \vert y)$, as discrete distributions with probabilities
\begin{equation} \label{eq:categorical_distribution}
    P_k = \frac{\exp \alpha_k}{\sum_{i=1}^K\exp \alpha_i },
\end{equation}
for $k \in \{1, \dots, K\}$, where $|\mathcal{U}|=K$ is a model parameter\footnote{$|\mathcal{U}|$ denotes the number of elements in the range of $U$.}. The unnormalized log-probabilities (\emph{logits}) $\alpha_k$ are either computed by ANNs as functions of the conditioning variable (i.e., $y$ for $q_{\boldsymbol{\zeta}}(u \vert y)$), where the parameters $\zeta$ represent the ANN weights, or treated as learnable parameters directly (i.e., $\boldsymbol{\xi}$ for $q_{\boldsymbol{\xi}}(u)$).

In order to formally introduce the encoder functions for marginal (i.e., $f_{m}(\cdot)$) and conditional (i.e., $f_{c}(\cdot)$) variants for a given set of parameters $\{\boldsymbol{\xi},
\boldsymbol{\phi},  \boldsymbol{\zeta}\}$, we first define the following sample losses for a given $X=x$ and $k=1, ..., K$, based on the loss functions in Eqs.~\eqref{eq:proposed_loss_marginal} and~\eqref{eq:proposed_loss_conditional}:
\begin{align}
\label{eq:l_m}
    \ell_{\mathrm{m}}(k, x) &= \mathbb{E}_{p(y \vert x)} \Big[-\log q_{\boldsymbol{\xi}}(k) + \lambda  \cdot  d(x, g_{\boldsymbol{\phi}}(k, y))\Big], \\ \label{eq:l_c}
    \ell_{\mathrm{c}}(k, x) &= \mathbb{E}_{p(y \vert x)} \Big[-\log q_{\boldsymbol{\zeta}}(k \vert y) + \lambda  \cdot  d(x, g_{\boldsymbol{\phi}}(k, y))\Big].
\end{align} Note that although the encoder does not have access to the realization of the side information $Y=y$ or the joint distribution $p(x,y)$, we assume that it can get i.i.d. samples from the conditional distribution $Y \sim p(y \vert x)$. We highlight that this is weaker than assumptions in Theorem~\ref{theo:WZ}, where it is assumed that the joint distribution $p(x,y)$ is known to both the encoder and the decoder. Since the conditional expectations in Eqs.~\eqref{eq:l_m} and~\eqref{eq:l_c}, which depend on $p(y \vert x)$, are not known in closed form, both encoder formulations will rely on sample means to approximate them as the following:
\begin{alignat}{1}
    \label{eq:l_m_2}
      \ell_{\mathrm{m}}(k, x) &\approx \frac{1}{N} \sum_{\{y_n \sim p(y \vert x) \, | \, 1 \leq n \leq N\}} \Big[-\log q_{\boldsymbol{\xi}}(k) + \lambda  \cdot  d(x, g_{\boldsymbol{\phi}}(k, y_n))\Big], \\ \label{eq:l_c_2}
      \ell_{\mathrm{c}}(k, x) &\approx \frac{1}{N} \sum_{\{y_n \sim p(y \vert x) \, | \, 1 \leq n \leq N\}} \Big[-\log q_{\boldsymbol{\zeta}}(k \vert y_n) + \lambda  \cdot  d(x, g_{\boldsymbol{\phi}}(k, y_n))\Big].
\end{alignat} Note that the sample means in Eqs.~\eqref{eq:l_m_2} and~\eqref{eq:l_c_2} are unbiased estimators of the conditional expectations in Eqs.~\eqref{eq:l_m} and~\eqref{eq:l_c}, based on averaging over $N$ samples obtained from the conditional distribution $y_{n} \sim p(y \vert x)$. Next, the encoder functions (see Section~\ref{sec:optimal_operational_rd_function} for the relevant definitions) for marginal and conditional variants are set as:
\begin{align}\label{eq:f_m}
    f_{\mathrm{m}}(x) &= \argmin_{k \in K} \ell_{\mathrm{m}}(k, x), \\ \label{eq:f_c}
    f_{\mathrm{c}}(x) &= \argmin_{k \in K} \ell_{\mathrm{c}}(k, x).
\end{align} Unlike the NTC framework (see Section~\ref{sec:ntc_intro})
that relies on an ordered latent space in $\mathbb{R}^{n_l}$, as the \emph{image} of $\mathbb{R}^{n_s}$ under the encoder (\emph{analysis}) function, we now explicitly enumerate the codebook vectors in our new formulation. Remark that the encoder functions in Eqs.~\eqref{eq:f_m} and~\eqref{eq:f_c} do not have any learnable parameters of their own, but depend on the probability models and the decoding function that are parameterized by $\{\boldsymbol{\xi}, \boldsymbol{\zeta}\}$ and $\{\boldsymbol{\phi}\}$, respectively. They also operate on an unordered categorical latent space, where $k$ denotes the category of the quantization index. Using the encoder functions in Eqs.~\eqref{eq:f_m} and~\eqref{eq:f_c}, we can express Eqs.~\eqref{eq:proposed_loss_marginal} and~\eqref{eq:proposed_loss_conditional} equivalently as:
\begin{alignat}{2}
    \label{eq:L_m}
    L_\mathrm{m}(\boldsymbol{\xi}, \boldsymbol{\phi}) &=  \mathbb{E}_{p(x)} \; \ell_{\mathrm{m}}(f_{\mathrm{m}}(x), x) &&= \mathbb{E}_{p(x)} \min_{k \in K} \ell_{\mathrm{m}}(k,x), \\ \label{eq:L_c}
    L_\mathrm{c}(\boldsymbol{\zeta}, \boldsymbol{\phi}) &=  \mathbb{E}_{p(x)} \; \ell_{\mathrm{c}}(f_{\mathrm{c}}(x), x) &&= \mathbb{E}_{p(x)} \min_{k \in K} \ell_{\mathrm{c}}(k,x).
\end{alignat}

The learnable parameters in loss functions $ L_\mathrm{m}(\boldsymbol{\xi}, \boldsymbol{\phi})$ and $ L_\mathrm{c}(\boldsymbol{\zeta}, \boldsymbol{\phi})$ can be jointly optimized with stochastic gradient descent (SGD) since the loss functions are differentiable with respect to them. We can compute the gradients using automatic differentiation methods, as implemented in deep learning frameworks such as JAX~\cite{jax}. By varying the trade-off parameter $\lambda$, we obtain different points in the achievable R-D region. Here, SGD replaces the expectations in the loss functions by averages over batches of $B$ from $p(x,y)$ samples, and exchanges the order of differentiation and summation, due to linearity. For example, considering a sample loss $\ell_{\boldsymbol{\mu}}(x,y)$ with parameters $\boldsymbol \mu$ (represented as one of the sample loss functions inside the brackets in Eqs.~\eqref{eq:proposed_loss_marginal} or \eqref{eq:proposed_loss_conditional}) we approximate:
\begin{align} \label{eq:sgd_approx}
    \frac{\partial }{\partial \boldsymbol{\mu} }\mathbb{E} [\;  \ell_{\boldsymbol{\mu}}(x,y)) \; ] \approx \frac{1}{|B|} \sum_{(x, y)\in B} \frac{\partial \ell_{\boldsymbol{\mu}}(x,y)}{\partial \boldsymbol{\mu}} \; . 
\end{align} 
We remark that this is, again, an unbiased estimator of the derivative of the expectation, based on averaging the derivatives of $\ell_{\boldsymbol{\mu}}$ over a batch of $B$ source samples.

The design choices discussed in this section keep the parametric families as general as possible and do not unnecessarily impose any structure. Specifically, this allows the encoder to learn, if needed, quantization schemes that involve discontiguous bins in the source space, akin to the \emph{random binning} operation in the achievability part of the WZ theorem~\cite{Wyner:IT:76}, and resembling the systematic partitioning of the quantized source space with cosets in DISCUS~\cite{DISCUS}.

\subsection{Induced Operational Schemes} \label{subsec:operational_schemes}

\begin{figure*}[t]
\centering
\begin{subfigure}[b]{.7\columnwidth}
   \includegraphics[width=1\linewidth]{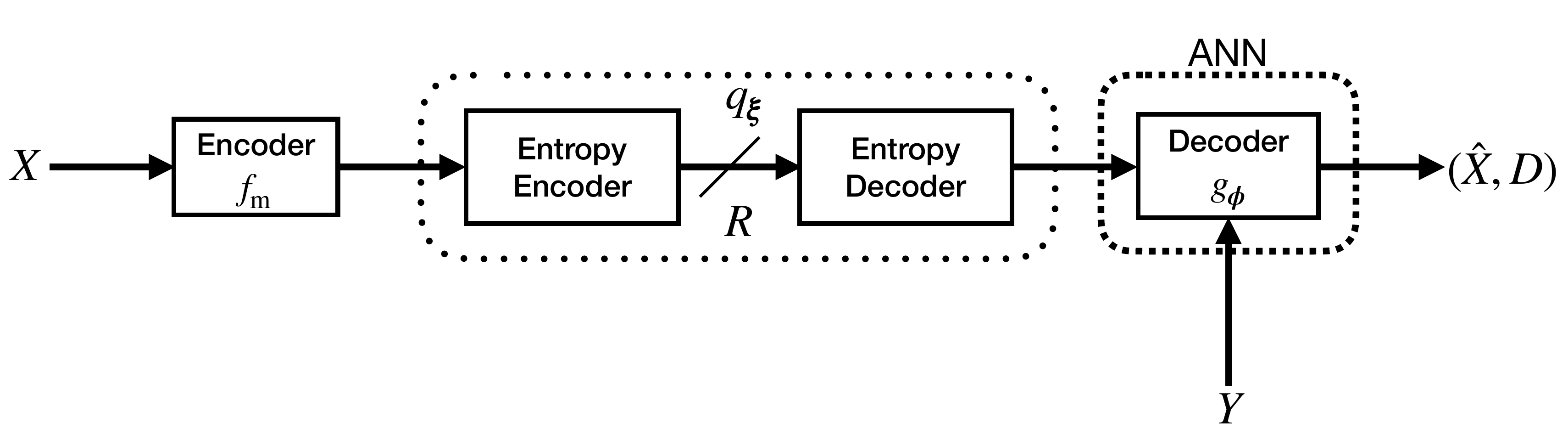}
   \caption{}
   \label{fig:sys_marginal} 
\end{subfigure}
\hfill
\begin{subfigure}[b]{.7\columnwidth}
   \includegraphics[width=1\linewidth]{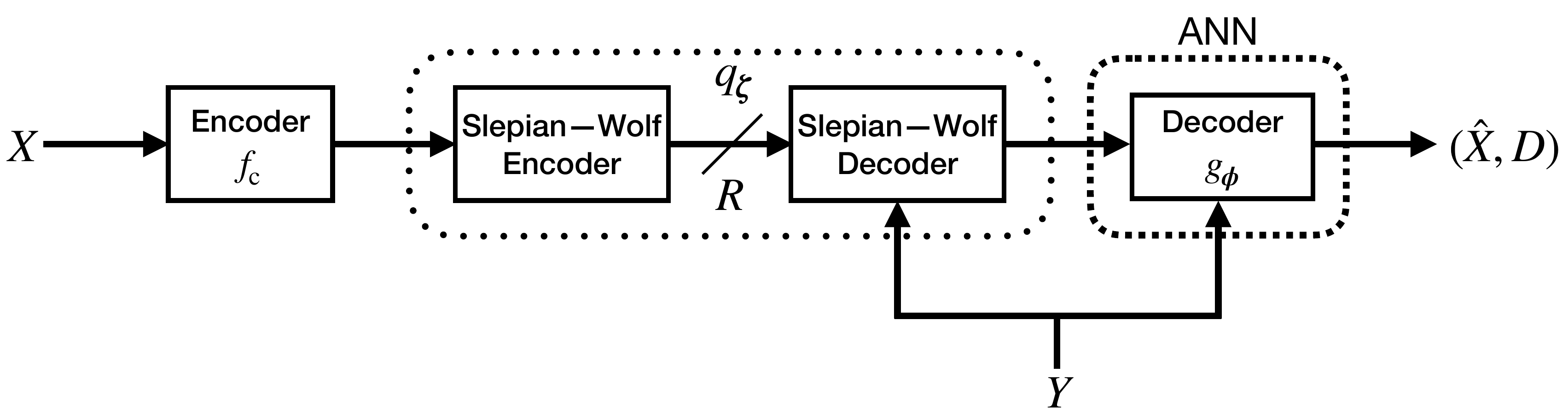}
   \caption{}
   \label{fig:sys_conditional}
\end{subfigure}
\caption{The two lossy compression systems that we consider, based on upper bounds in Eqs.~\eqref{eq:upper_bound_marg_2} and~\eqref{eq:upper_bound_cond_2}: learned compressor using a classic entropy coder (a), and learned compressor using an ideal Slepian--Wolf coder (b). See Section~\ref{subsec:operational_schemes} for the relevant discussion.}
\label{fig:sys}
\end{figure*} 

The two variants of the objective functions we formulated in Eqs.~\eqref{eq:proposed_loss_marginal} and~\eqref{eq:proposed_loss_conditional} directly correspond to two distinct solutions to the WZ problem (see Fig.~\ref{fig:sys}). By the marginal formulation, we handle the quantization and binning parts jointly and assume a classic entropy coder (see Fig.~\ref{fig:sys_marginal}). In the conditional case, we instead take a two-step approach by having an ideal SWC that is coupled with a learned compressor, as seen in Fig.~\ref{fig:sys_conditional}. We also inherit some standard design assumptions in our formulations, from both ECSQ~\cite{Gyorgy} and the popular class of
NTC-based neural compressors~\cite{balle_journal} (see Section~\ref{sec:relevant_quantizer_designs} for the relevant discussion). Although we compress each source realization one at a time, as in we do \emph{one-shot} quantization, we assume that the entropy coding part is carried out over large blocks of the quantized source elements, i.e., in a \emph{high-order} fashion~\cite{Gish}. As such, the upper bound in Eq.~\eqref{eq:upper_bound_marg_2} corresponds to a compression rate of a system employing a one-shot encoder and a classic entropy code which asymptotically achieves a rate equal to the cross-entropy $\mathbb{E}_{p(x)}[-\log q_{\boldsymbol{\xi}}(u)]_{u=f_{\mathrm{m}}(x)}$. This choice is justified, as it has been shown that the actual rates achievable by a properly designed entropy code are only negligibly above the entropy values~\cite{universal_coding}. Similarly, the upper bound in Eq.~\eqref{eq:upper_bound_cond_2} coincides with a compression rate of a design having a one-shot compressor and an ideal SW entropy coder which asymptotically achieves the cross-entropy $\mathbb{E}_{p(x, y)}[-\log q_{\boldsymbol{\zeta}}(u \vert y)]_{u=f_{\mathrm{c}}(x)}$. The ideal SW entropy coder~\cite{Slepian:IT:73}, which achieves a compression rate of $H(U \vert Y)=H(U) - I(U;Y)$, performs entropy coding and also further exploits the correlation between $U$ and $Y$ at the same time. Some practical SW coding schemes (e.g.,~\cite{ldpc_0, SW_0, SW_1}) are syndrome-based ones with linear channel codes, which exploit the statistical correlation between $U$ and $Y$ but abandon entropy coding. As mentioned in~\cite{ecsq_w_side_info}, the achievable rate of these practical SW schemes can be upper bounded by $\log(A) - I(U;Y)$, where $A$ denotes the fixed number of partitions allowed in the quantizer. Therefore, assigning $H(U \vert Y)$ and $\log(A) - I(U;Y)$ as the compression rate yields lower and upper bounds on the operational achievable rate by an SWC, respectively. The two variants in~\cite{ecsq_w_side_info}, assuming either an ideal or a practical SWC, will be reported later in Section~\ref{subsec:quadratic_Gaussian} (more specifically, in Fig.~\ref{fig:y=x+n_var_n=0.1}).

\section{Experimental Setup}
\label{sec:experimental_setup}

The asymptotic WZ formula in Eq.~\eqref{eq:WZ} has a closed-form expression only in a few special cases. To evaluate how close our neural bounds, presented in Section~\ref{sec:method}, can get to the known R-D function, we first consider the following correlation model: let $X$ and $Y$ be correlated, zero mean and i.i.d. Gaussian memoryless sources, and let the distortion metric be MSE. Then, the WZ R-D function is
\begin{equation} \label{eq:WZ_CD}
    R_{\text{WZ}}(D) =  \frac{1}{2}\log\left( \frac{\sigma_{x \vert y}^{2}}{D} \right), \; \; 0 \leq D \leq \sigma_{x \vert y}^{2},
\end{equation}
where $ \sigma_{x \vert y}^{2}$ denotes the conditional variance of $X$ given $Y$. For correlation structures of $X=Y+N$ and $Y=X+N$ with $N \sim \mathrm{N}(0,\sigma_{n}^2)$, we have $\sigma_{x \vert y}^2 = \sigma_{n}^2$ and $\sigma_{x \vert y}^2 = \frac{\sigma_{x}^{2}\sigma_{n}^2}{\sigma_{x}^{2} + \sigma_{n}^2}$, respectively.

Next, we consider the source model presented previously in Section~\ref{sec:ntc_intro}: let $X \sim \mathrm{Laplace}(0;1)$, $Y= \mathrm{sgn}(X)$, and the distortion criterion be MSE. Remarking that in this case the encoder only needs to communicate $|X| \sim \mathrm{Exponential}(1)$, we can compare the R-D performance achieved by our compressor with that of optimal ECSQ having Laplacian or exponential distributions as its input, whose performances are already characterized by Sullivan~\cite{sullivan}. Note that the optimal ECSQ    performance of Laplacian corresponds to the entropy--distortion function of $X$ (Definition~\ref{def:E_D}) and that of exponential corresponds to the entropy--distortion function of $X$ given $Y$ (Definition~\ref{def:entropy_opt_side_info}).) Although this may appear as a simple example of a deterministic correlation model between $X$ and $Y$, as we will see, it will serve to illustrate a case of binning mechanism through which our proposed model outperforms the NTC-based neural compressor, by exhibiting interpretable mapping functions in the source space (which will be shown in Section~\ref{subsec:laplacian}). 

Note that in spite of considering Gaussian or Laplacian sources, we do \emph{not} make any assumptions on the distribution of information sources in our formulations of the models. The parameters  $\{\boldsymbol{\xi},
\boldsymbol{\phi},  \boldsymbol{\zeta}\}$ are learned solely in a data-driven way from realizations of the sources, through the proposed loss functions in Eqs.~\eqref{eq:L_m} and \eqref{eq:L_c}.

For the conditional probabilistic model, $q_{\boldsymbol{\zeta}}$, and the decoding function, $g_{\boldsymbol{\phi}}$, we employ ANNs of three dense layers, with 100 units each (excluding the last one), and leaky rectified linear units as activation functions (again, excluding the last) for each of the layers. For the marginal probabilistic model $q_{\boldsymbol{\xi}}$, we directly have learnable parameters. In our experiments, we found that larger networks or different activation functions did not improve the results. The decoding function receives a concatenated vector of both its inputs, $u$ and $y$. We use Adam~\cite{adam}, a popular variant of SGD, and conduct our experiments using the JAX~\cite{jax} framework. We train all learning-based models for $100$ epochs, using randomly initialized network weights. We initially set the number of available indices as $K=32$, where $k \in \{1, \dots, K\}$ in Eq.~\eqref{eq:categorical_distribution}. We use a learning rate of $1 \times 10^{-3}$ and a batch size of $B=512$ (as in Eq.~\eqref{eq:sgd_approx}), which coincides with number of realizations sampled from the aforementioned correlation models.

VQ-based algorithms are known to be sensitive to initialization choices~\cite{initialization}. One simple method that is known to work well in practice is to initialize the codebook to random samples of the source. Since our codebook is implicitly represented in the decoder function, we emulate this initialization scheme by pretraining the decoder for 30 epochs. Specifically, we draw $2K$ i.i.d. samples $x$ from the source, and the same number of quantization indices $k$ from a uniform distribution. Pretraining then consists of minimizing the conditional expectation of \emph{only} the distortion terms of $\ell_\mathrm{m}$ or $\ell_\mathrm{c}$ in Eqs.~\eqref{eq:l_m_2} and~\eqref{eq:l_c_2}.

For the last $10$ epochs of the training step, we reduce the learning rate to $1 \times 10^{-4}$, as is standard practice in stochastic optimization. At the encoder side, we use a sample mean estimation by averaging over $N=24$ samples as in Eqs.~\eqref{eq:l_m_2} and~\eqref{eq:l_c_2}. We obtain all reported empirical estimates of rate and distortion values by averaging over $2^{20}$ source realizations. We will make our source code publicly available shortly.

\section{Experimental Results and Discussion}
\label{sec:discussion}

We present empirical results obtained for the source distributions explained in Section~\ref{sec:experimental_setup}. Gaussian and Laplacian experiments are discussed in Sections~\ref{subsec:quadratic_Gaussian} and~\ref{subsec:laplacian}, respectively. We conclude by analyzing connections to related work and elaborating on some avenues for future work in Section~\ref{subsec:discussion}.

\subsection{Gaussian Experiments}
\label{subsec:quadratic_Gaussian}
We consider both correlation models of $X=Y+N$ and $Y=X+N$, where we vary the quality level of the side information by altering the variance of the noise such that $\sigma_{n}^2 \in \{10^{-2}, 10^{-1} \}$. 

\begin{figure*}[t]
\centering
\begin{subfigure}{.5\textwidth}
  \centering
  \includegraphics[width=0.95\linewidth]{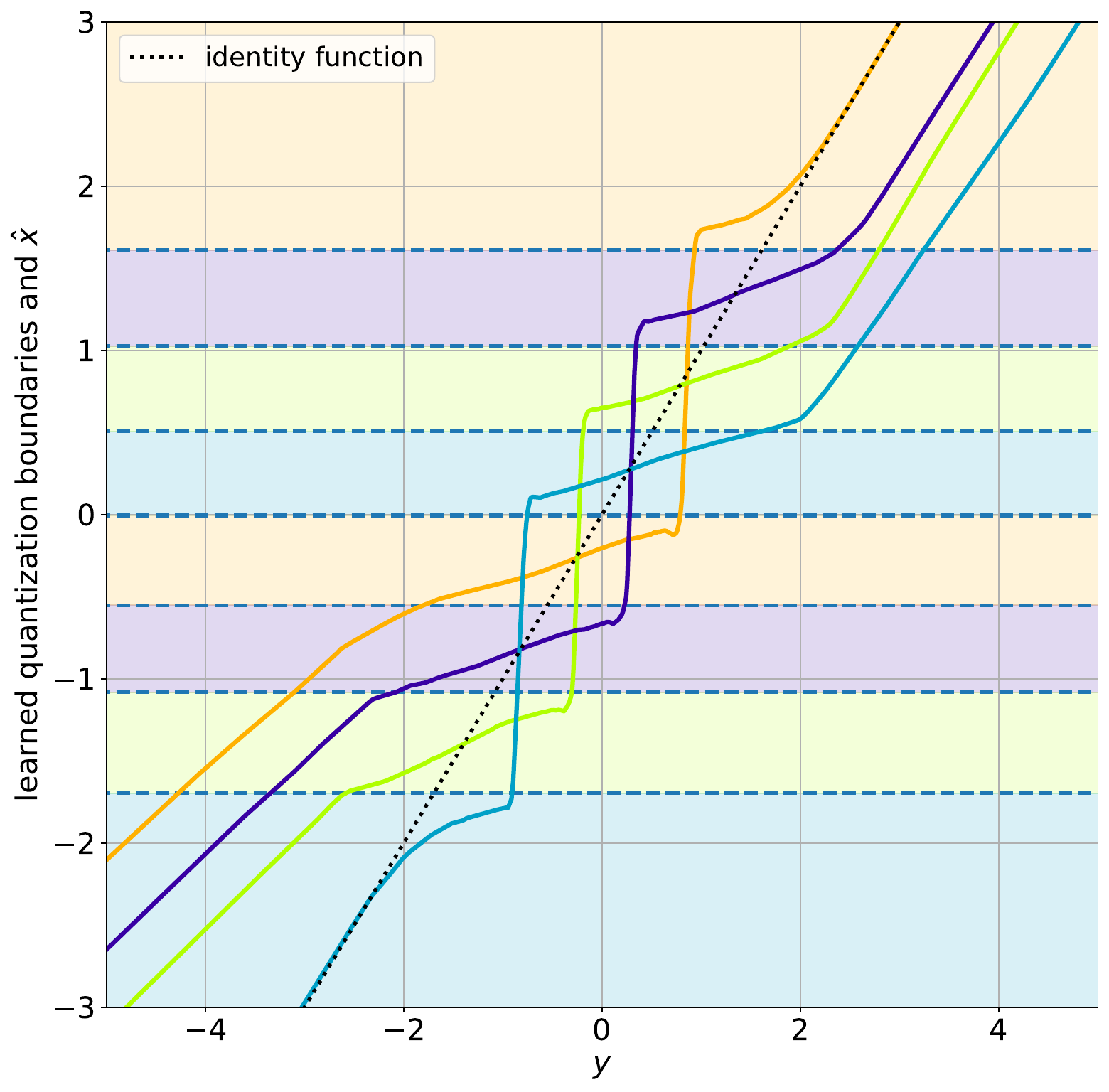}
  \caption{\footnotesize{$X=Y+N$ with $Y \sim N(0,1)$ and $\mathrm{N} \sim N(0,10^{-1})$.}}
  \label{fig:visualization_gaussian_1}
\end{subfigure}%
\hfill
\begin{subfigure}{.5\textwidth}
  \centering
  \includegraphics[width=0.98\linewidth]{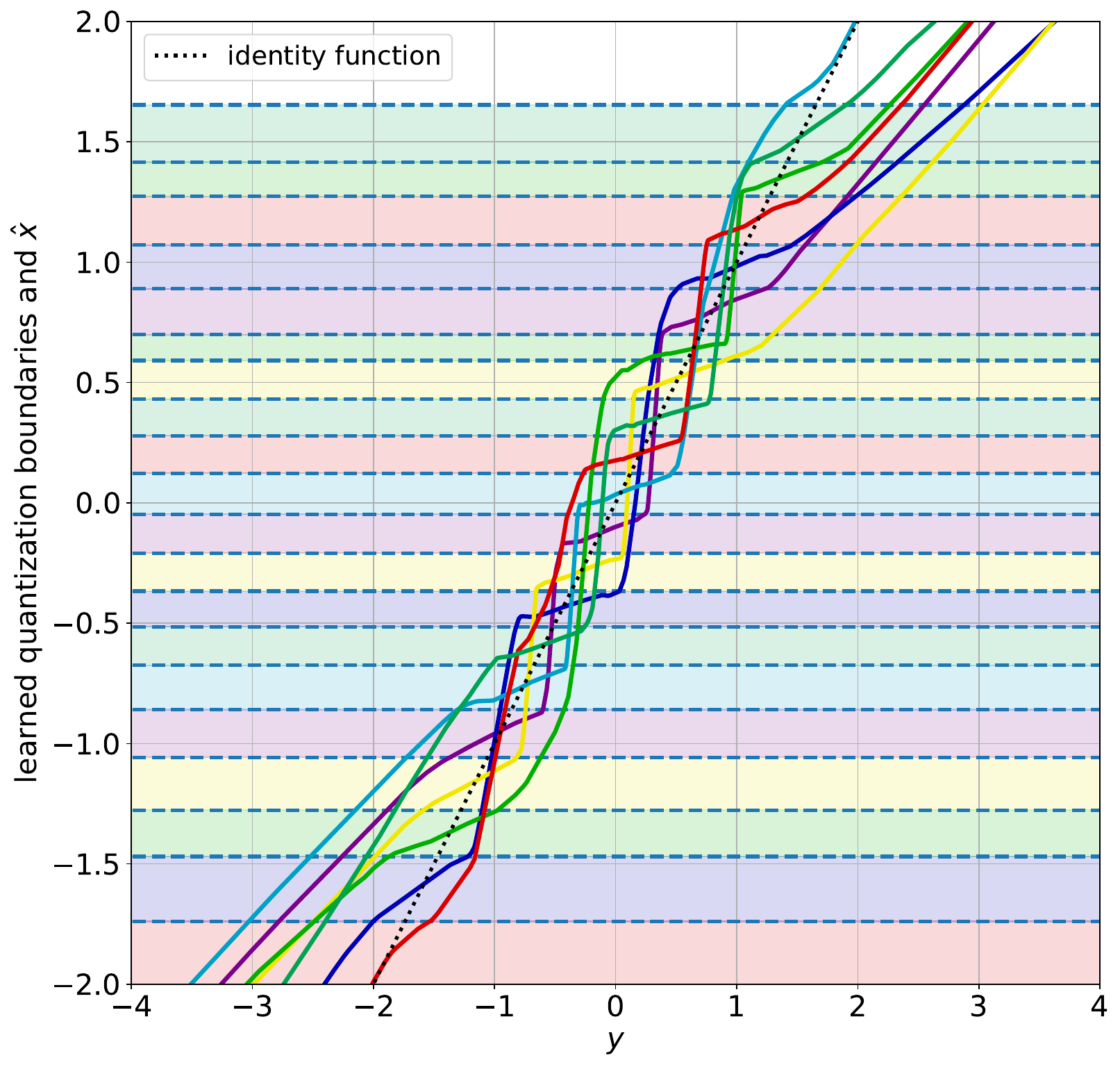}
  \caption{\footnotesize{$Y=X+N$ with $Y \sim N(0,1)$ and $\mathrm{N} \sim N(0,10^{-2})$.}}
  \label{fig:visualization_gaussian_2}
\end{subfigure}
\caption{Visualizations (best viewed in color) of the learned encoder $u =  \argmin_{k} \ell_{\mathrm{m}}(k, x)$ (see Eq.~\eqref{eq:f_m}) and neural decoder $\hat{x} = g_{\boldsymbol{\phi}}(u,y)$ of the marginal formulation (see Eq.~\eqref{eq:L_m}), for the Gaussian WZ setup. The dashed horizontal lines are quantization boundaries, and the colors between boundaries represent unique values of $u$. We depict the decoding function as separate plots for each value of $u$, using the same color assignment. The visualized models on the left and right panels achieve $-15.44$ dB at $2.00$ bits and $-25.67$ dB at $2.78$ bits, respectively. Note that both of the visualized models achieve a rate--distortion (R-D) performance better than the point-to-point R-D function, due to many-to-one mapping functions recovered by the proposed solution.}
\label{fig:visualizations_gaussian}
\end{figure*}

We first evaluate the marginal variant, whose system model and objective function are provided in Fig.~\ref{fig:sys_marginal} and Eq.~\eqref{eq:L_m}, respectively. In both panels of Fig.~\ref{fig:visualizations_gaussian}, we visualize the learned compressors obtained with this formulation. We remark that the learned compressors exhibit periodic grouping, binning-like behavior with respect to the source space, although no explicit structure was imposed onto the model architecture. Color coding of the bin indices reveals discontiguous quantization bins being learned. This demonstrates that learning-based methods are indeed capable of recovering very similar solutions to some of the handcrafted frameworks proposed for the WZ problem, such as the seminal DISCUS framework~\cite{DISCUS}. Note that this behavior is also analogous to the random binning procedure in the achievability part of the WZ theorem~\cite{Wyner:IT:76} (see Theorem~\ref{theo:WZ}).

The figure also shows that the learned compressors exhibit optimal decoder behavior within each quantization index. In the given setup, the optimal decoder disambiguates the quantization index from the received bin index $u$ (essentially undoing the binning or grouping), and reconstructs the source as~\cite{zamir},
\begin{equation} \label{eq:opt_decoder}
    \hat{x} = (1-\beta) \cdot y + \beta \cdot M(u), \;  \text{where} \; \beta \propto \sigma_{n}^2 \: ,
\end{equation}
where $M(\cdot)$ denotes the disambiguation procedure. The slopes of the learned curves are also sensitive to $\sigma_{n}^2$, as is evident from comparing both panels of Fig.~\ref{fig:visualizations_gaussian}.

We explain the behavior of this learned model as follows. The encoder quantizes the source and subsequently bins the quantization index using the learned joint statistics of $f_{m}(X)$ and $Y$, where $f_{m}(\cdot)$ refers to the encoding function, yielding $u$ (see Eq.~\eqref{eq:f_m}). Note that the encoder does \emph{not} explicitly have access to the realization $Y=y$. The decoder then disambiguates the received bin index and deduces the quantization index, with the help of the side information. It subsequently estimates the source as $\hat{x}$, yielding the linear decoding functions within each quantization index with respect to the matching curve shown in both panels of Fig.~\ref{fig:visualizations_gaussian}.  Observe that the corresponding decoding functions in matching quantization indices exhibit kinks close to the quantization boundaries. This demonstrates that the model indeed tries to adopt linear functions, as is the optimal decoder behavior in Eq.~\eqref{eq:opt_decoder}.

Looking at the encoders learned by the conditional formulation, whose system model and objective function are provided in Fig.~\ref{fig:sys_conditional} and Eq.~\eqref{eq:L_c} respectively, we find no evidence of binning occurring in the source space (not depicted). We argue that this is due to having an ideal SWC in its system design instead of a classic entropy coder as in the marginal variant (see Fig.~\ref{fig:sys_marginal}). We speculate that this choice of entropy coding scheme instead encourages the model to leave the task of binning solely to the ideal SW code, rather than to the quantizer. The SW code may make use of a high dimensional channel code (e.g., as in DISCUS~\cite{DISCUS}) that enables the model to essentially execute binning over long sequences, i.e., in a multi-shot fashion. Note that such a high-order binning scheme is much more efficient than the one that could be achieved by the quantizer, which could only bin the induced quantization indices in a one-shot manner, since it compresses each source realization one at a time.

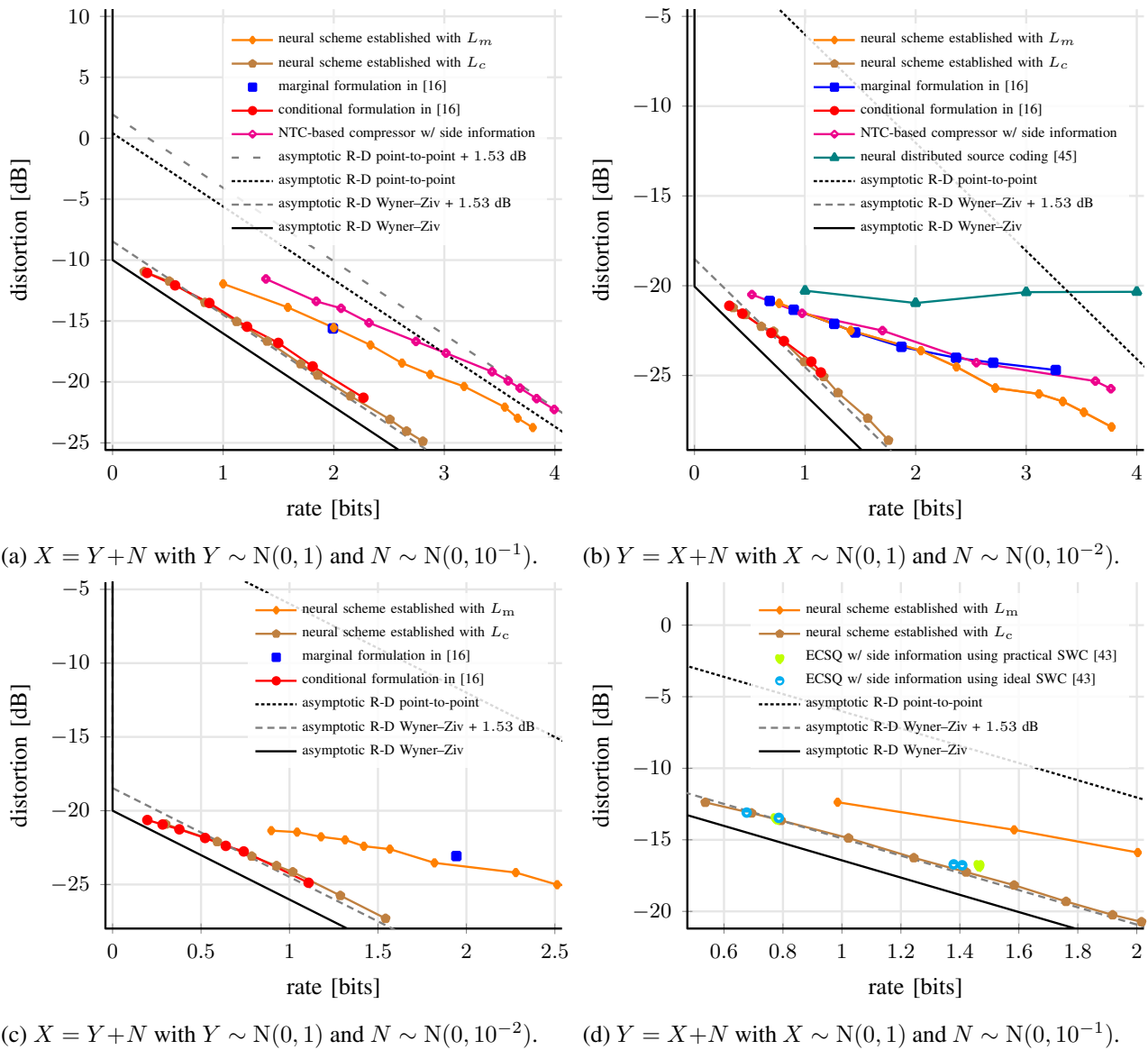
\begin{figure*}
\begin{subfigure}{.48\columnwidth}
    \raggedleft
    \begin{tikzpicture}[trim axis right]
    \begin{axis}[
      height=.28\textheight,
      width=.85\linewidth,
      scale only axis,
      xlabel={rate [bits]},
      ylabel={distortion [dB]},
      xmin=0.,
      xmax=4.0,
      ymin=-25.,
      ymax=10.,
      legend pos=north east,
      legend style={font=\tiny},
      ]
      \addplot[color=orange, mark=diamond*] table{
     0.99991536 -11.946896314620972
     1.5850284 -13.883405923843384
     1.9999739 -15.552754402160645
     2.3323302 -16.972007751464844
     2.6175106 -18.454054594039917
    2.8724792 -19.391801357269287
      3.1790128 -20.37123203277588
    3.5471265 -22.068610191345215
    3.6638668 -22.976665496826172
    3.800669 -23.74577045440674
      };
      \addplot[color=brown, mark=pentagon*] table{
    0.2863052 -10.949747562408447
    0.513098 -11.737004518508911
    0.8352904 -13.48630428314209
    1.1222035 -15.047709941864014
    1.3991282 -16.65987730026245
    1.7008452 -18.51858139038086
    1.8545041 -19.44868803024292
    2.1517706 -21.156933307647705
    2.5079832 -23.08281898498535
    2.658597 -24.044439792633057
    2.8061166 -24.876220226287842
      };
       \addplot[color=blue,mark=square*,only marks] table {
        1.992392 -15.610814094543457
      };
      \addplot[color=red,mark=*] table {
        0.31238356 -11.05667233467102
        0.5654873 -12.075015306472778
        0.87530893 -13.522629737854004
        1.2161504 -15.470525026321411
        1.4988127 -16.797856092453003
        1.8096685 -18.724064826965332
        2.2686293 -21.29608154296875
      };
      \addplot[color=magenta, mark=halfsquare*] table {
        1.3866295 -11.548492908477783
        1.841862 -13.387084007263184
        2.0665174 -13.957363367080688
        2.3194141 -15.148096084594727
        2.7438128 -16.67616367340088
       3.0152524 -17.634437084197998
        3.432304 -19.162425994873047
        3.5775576 -19.914076328277588
        3.684813 -20.506529808044434
        3.8349023 -21.369247436523438
        3.9942348 -22.261009216308594
      };
      \addplot[color=gray, loosely dashed] table {
      0. 1.9468578937196708
      6. -34.17674158595807
      };
      \addplot[color=black,densely dotted] table {
        0. 0.41392685
        6. -35.70967263
      };
      \addplot[color=gray,densely dashed] table {
        0. -8.46706895786258
        6. -44.59066843786258
      };
      \addplot[color=black] table {
        0. 13.
        0. -10.
        6. -46.12359948
      };
      \addplot[color=orange, mark=diamond*] table{
        1.9999739 -15.552754402160645
      };
      \legend{neural scheme established with $L_m$, neural scheme established with $L_c$, marginal formulation in~\cite{ozyilkan2023learned}, conditional formulation in~\cite{ozyilkan2023learned}, NTC-based compressor w/ side information, asymptotic R-D point-to-point + $1.53$ dB, asymptotic R-D point-to-point, asymptotic R-D Wyner--Ziv + $1.53$ dB, asymptotic R-D Wyner--Ziv};
    \end{axis}
    \end{tikzpicture}
  \caption{$X=Y+N$ with $Y \sim \mathrm{N}(0,1)$ and $N \sim \mathrm{N}(0,10^{-1})$.}
  \label{fig:x=y+n_var_n=0.1}
\end{subfigure}%
\hfill%
\begin{subfigure}{.48\columnwidth}
    \raggedleft
    \begin{tikzpicture}[trim axis right]
    \begin{axis}[
      height=.28\textheight,
      width=.85\linewidth,
      scale only axis,
      xlabel={rate [bits]},
      ylabel={distortion [dB]},
      xmin=0.,
      xmax=4.0,
      ymin=-28.75,
      ymax=-5.,
      legend pos=north east,
      legend style={font=\tiny},
      ]
      \addplot[color=orange,mark=diamond*] table {
        0.7656008 -20.981309413909912
        1.4131868 -22.49509572982788
        2.0475862 -23.61626625061035
        2.3694816 -24.518966674804688
        2.7210867 -25.69542646408081
        3.1154644 -26.024339199066162
        3.331149 -26.4467716217041
        3.5216622 -27.046027183532715
        3.7724364 -27.866063117980957
      };
      \addplot[color=brown, mark=pentagon*] table{
      0.3497216 -21.22068166732788
      0.4586787 -21.610662937164307
      0.60386175 -22.268359661102295
      0.7156547 -22.541887760162354
      0.99397546 -24.235875606536865
      1.1690772 -25.064470767974854
      1.298243 -25.959668159484863
      1.5664442 -27.382283210754395
      1.7538832 -28.610849380493164
      };
     \addplot[color=blue,mark=square*] table {
        0.679996 -20.8489727973938
        0.89723724 -21.337780952453613
        1.2638619 -22.123868465423584
        1.4534979 -22.6043963432312
        1.8716173 -23.40317964553833
        2.3653505 -24.013419151306152
        2.7019472 -24.288642406463623
        3.2669206 -24.69569683074951
      };
      \addplot[color=red,mark=*] table {
        0.31479982 -21.115736961364746
        0.43293902 -21.552035808563232
        0.6946749 -22.629497051239014
        0.80841285 -23.079593181610107
        1.0567338 -24.226911067962646
        1.1426133 -24.82905149459839
      };
       \addplot[color=magenta, mark=halfsquare*] table {
         0.51899236 -20.492868423461914
         0.97265935 -21.535344123840332
         1.7 -22.5
         2.5490305 -24.28478717803955
         3.6254618 -25.311644077301025
          3.765881 -25.736722946166992
         };
     \addplot[color=teal,mark=triangle*, mark size=2pt] table {
        1. -20.27697580250686
        2. -20.964133926201463
        3. -20.357155522665344
        4. -20.3388973674
      };
      \addplot[color=black,densely dotted] table {
        0. 0.0
        6. -36.12359947967774
      };
      \addplot[color=gray,densely dashed] table {
        0. -18.510282695689007
        6. -54.63388217536675
      };
      \addplot[color=black] table {
        0. 10.
        0. -20.043213737826427
        6. -56.16681321750417
      };
       \addplot[color=orange,mark=diamond*] table {
        0.7656008 -20.981309413909912
        1.4131868 -22.49509572982788
        2.0475862 -23.61626625061035
        2.3694816 -24.518966674804688
        2.7210867 -25.69542646408081
        3.1154644 -26.024339199066162
        3.331149 -26.4467716217041
        3.5216622 -27.046027183532715
        3.7724364 -27.866063117980957
      };
      \legend{neural scheme established with $L_m$, neural scheme established with $L_c$, marginal formulation in~\cite{ozyilkan2023learned}, conditional formulation in~\cite{ozyilkan2023learned}, NTC-based compressor w/ side information, neural distributed source coding~\cite{NDSC}, asymptotic R-D point-to-point, asymptotic R-D Wyner--Ziv + $1.53$ dB, asymptotic R-D Wyner--Ziv};
    \end{axis}
    \end{tikzpicture}
    \caption{$Y=X+N$ with $X \sim \mathrm{N}(0,1)$ and $N \sim \mathrm{N}(0,10^{-2})$.}
    \label{fig:y=x+n_var_n=0.01}
\end{subfigure}%
\vfill
\begin{subfigure}{.48\columnwidth}
    \raggedleft
    \begin{tikzpicture}[trim axis right]
     \begin{axis}[
      height=.22\textheight,
      width=.85\linewidth,
      scale only axis,
      xlabel={rate [bits]},
      ylabel={distortion [dB]},
      xmin=0.,
      xmax=2.5,
      ymin=-27.5,
      ymax=-5.,
      legend pos=north east,
      legend style={font=\tiny},
      ]
    \addplot[color=orange, mark=diamond*] table{
   0.8970345 -21.3498592376709
   1.0428817 -21.45085334777832
   1.1783129 -21.76849126815796
  1.314236 -21.97446346282959
  1.4203601 -22.40572690963745
   1.5675151 -22.602667808532715
   1.8194658 -23.5348629951477056
   2.2792258 -24.1913104057312
   2.5130181 -25.01333713531494
   2.7386355 -25.20796298980713
      };
      \addplot[color=brown, mark=pentagon*] table{
        0.29995692 -20.909433364868164
        0.5915958 -22.098681926727295
        0.78665215 -23.08292865753174
        0.926544 -23.72751235961914
        1.0185641 -24.152238368988037
        1.2874504 -25.751450061798096
        1.5429667 -27.302775382995605
       };
      \addplot[color=blue,mark=square*,only marks] table {
        1.9432522 -23.07220697402954
      };
      \addplot[color=red,mark=*] table {
        0.1964734 -20.630433559417725
        0.28251672 -20.933425426483154
        0.37739894 -21.26171350479126
        0.52286476 -21.847405433654785
        0.63997626 -22.381272315979004
        0.741162 -22.759950160980225
        1.1078129 -24.88246202468872
      };
      \addplot[color=black,densely dotted] table {
        0. 0.04321374
        6. -36.08038574
      };
      \addplot[color=gray,densely dashed] table {
        0. -4.
        0. -18.46706895786258
        6. -54.59066843786258
      };
      \addplot[color=black] table {
        0. -4.
        0. -20.
        6. -56.12359948
      };
      \legend{neural scheme established with $L_\mathrm{m}$, neural scheme established with $L_\mathrm{c}$, marginal formulation in~\cite{ozyilkan2023learned}, conditional formulation in~\cite{ozyilkan2023learned}, asymptotic R-D point-to-point, asymptotic R-D Wyner--Ziv + $1.53$ dB, asymptotic R-D Wyner--Ziv};
    \end{axis}
    \end{tikzpicture}
    \caption{$X=Y+N$ with $Y \sim \mathrm{N}(0,1)$ and $N \sim \mathrm{N}(0,10^{-2})$.}
    \label{fig:x=y+n_var_n=0.01}
\end{subfigure}
\hfill
\begin{subfigure}{.48\columnwidth}
    \raggedleft
    \begin{tikzpicture}[trim axis right]
    \begin{axis}[
      height=.22\textheight,
      width=.85\linewidth,
      scale only axis,
      xlabel={rate [bits]},
      ylabel={distortion [dB]},
      xmin=0.5,
      xmax=2.0,
      ymin=-20.7,
      ymax=2.5,
      legend pos = north east,
      legend style={font=\tiny},
      ]
      \addplot[color=orange, mark=diamond*] table{
        0.9853333 -12.374297380447388
        1.5840037 -14.302208423614502
        2.0034254 -15.889244079589844
      };
      \addplot[color=brown, mark=pentagon*] table{
       0.5365629 -12.386820316314697
       0.6934582 -13.131182193756104
       0.7933945 -13.649369478225708
       1.0222838 -14.880499839782715
       1.2438215 -16.257129907608032
       1.4216777 -17.258265018463135
       1.5840105 -18.16953420639038
       1.7601674 -19.310877323150635
       1.9179085 -20.246894359588623
       2.0160575 -20.73216438293457
       };
      \addplot[color=lime, mark=heart, only marks] table{
      0.776 -13.4008379993
      0.781 -13.4678748622
      1.464 -16.6756154008
      1.466 -16.757175447
      };
      \addplot[color=cyan, mark=halfcircle*, only marks] table{
      0.677 -13.0980391997
      0.786 -13.4678748622
      1.379 -16.7162039656
      1.408 -16.7778070527
      };
      \addplot[color=black,densely dotted] table {
        0. 0.0
        6. -36.12359947967774
      };
      \addplot[color=gray,densely dashed] table {
        0. -8.88099580944483
        6. -45.00459528912258
      };
      \addplot[color=black] table {
        0. -10.41392685158225
        6. -46.53752633126
      };
      \legend{neural scheme established with $L_\mathrm{m}$, neural scheme established with $L_\mathrm{c}$, ECSQ w/ side information using practical SWC~\cite{ecsq_w_side_info}, ECSQ w/ side information using ideal SWC~\cite{ecsq_w_side_info}, asymptotic R-D point-to-point, asymptotic R-D Wyner--Ziv + $1.53$ dB, asymptotic R-D Wyner--Ziv};
    \end{axis}
\end{tikzpicture}
\caption{$Y=X+N$ with $X \sim \mathrm{N}(0,1)$ and $N \sim \mathrm{N}(0,10^{-1})$.}  \label{fig:y=x+n_var_n=0.1}
\end{subfigure}
\caption{Rate--distortion (R-D) performances obtained with marginal and conditional formulations, as in Eqs.~\eqref{eq:L_m} and \eqref{eq:L_c} respectively, NTC-based compressor with side information and related works in~\cite{NDSC, ecsq_w_side_info}. We also reproduce results from our previous work~\cite{ozyilkan2023learned}. We consider various Gaussian WZ setups with two different correlation structures, and plot the empirical results versus the asymptotic bounds. The $1.53$ dB distortion offset refers to the space-filling loss that the entropy-constrained one-shot lattice quantizer is subjected to in a high-rate regime.}
\label{fig:RD_results_gaussian}
\end{figure*}

In Fig.~\ref{fig:RD_results_gaussian}, we provide R-D performances obtained with these two different formulations under various set of Gaussian WZ settings. As seen in all panels of the figure, the marginal formulation yields a better performance compared to the point-to-point R-D function across all rate ranges. We argue that this is mainly due to the learned binning behavior (demonstrated in both panels of Fig.~\ref{fig:visualizations_gaussian}), resulting in rate reduction. However, this compressor does not reach the asymptotic WZ R-D bound provided in Eq.~\eqref{eq:WZ_CD}. In the figure, $1.53$ dB refers to the MSE gap that the entropy-constrained scalar (one-shot) lattice quantizer is subjected to in a high-rate regime \cite{quantization} with respect to point-to-point R-D function, due to space-filling loss (also known as cubic loss~\cite{cubic_loss}). As the compressor obtained through the marginal formulation compresses and consecutively bins each scalar input one by one, it is subjected both to the space-filling loss~\cite{high_resolution_quantization} during the quantization step, as well as to the loss coming from binning non-uniformly distributed quantization indices in a one-shot fashion. The achievability part of the WZ theorem, by comparison, considers binning of long sequences. This type of \emph{compress--bin}~\cite{network_info_theo} is much more efficient than the one-shot instance the marginal formulation is subjected to, as it exploits the Asymptotic Equipartition Property Theorem~\cite{elements_of_information_theory} that results in the asymptotic optimality of random binning. Note that visualizing the system behavior as in Fig.~\ref{fig:visualizations_gaussian} is difficult for more than one-dimensional $x$ and $y$ (i.e., for larger blocklengths).

Examining the empirical results obtained with the conditional formulation, we observe that its performance is indeed closer to the asymptotic WZ R-D bound. We explain the improved R-D performance of this variant as follows. When binning is left to the ideal SW code, the performance loss of such a learned WZ compressor only comes from the quantization part alone. This line of reasoning was also followed by the practical code design in~\cite{tcq_ldpc}, where the authors make use of a combination of a classic quantizer (without binning) and a powerful SW coding scheme, implemented with irregular low-density parity-check (LDPC) codes, in order to achieve the theoretical limit of $H(Q(X) \vert Y)$, where $Q(X)$ refers to the quantized source. Hence, minimizing $L_\mathrm{c}$ objective function in Eq.~\eqref{eq:L_c} corresponds to learning one-shot quantizer and dequantizer components, reducing the WZ problem to a SW problem with a data-driven quantization.

We also obtain R-D performances achieved by the NTC-based compressor~\cite{balle_journal} with side information (see Section~\ref{sec:ntc_intro} for the relevant discussion), and plot them in Figs.~\ref{fig:x=y+n_var_n=0.1} and~\ref{fig:y=x+n_var_n=0.01}. As seen, the NTC-based framework scores below point-to-point R-D function at lower rates while gradually achieving worse performance with respect to WZ R-D bound as rate gets higher. In Fig.~\ref{fig:NTC_vis} (in the Appendix), we provide the visualization of an NTC-based compressor with side information that achieves an R-D performance better than the point-to-point R-D bound. Surprisingly, color coding of its quantization bin indices unveils that at low rates, this neural compressor learns exactly symmetric groupings (around $x=0$) in the source space, unlike the periodic-like mappings seen in both panels of Fig.~\ref{fig:visualizations_gaussian}. We remark that the Figs.~\ref{fig:visualization_gaussian_1} and~\ref{fig:NTC_vis} (in the Appendix) consider the same Gaussian setup and also score comparable rate values while the NTC-based compressor performs about 1.5 dB worse in distortion. The superior performance of periodic-like mappings of our marginal formulation compared to symmetric ones of NTC is consistent with the key design intuition in DISCUS~\cite{DISCUS}, where the coset constructions in the quantization codebook space are done to keep the minimum distance between any two codewords within every coset as large as possible. We also observe that the learned decoders of NTC-based compressor adopt linear functions within each quantization index with respect to matching curve, which is, again, aligned with the optimal decoder behavior that carries out minimum mean square error estimation as in Eq.~\eqref{eq:opt_decoder}. These findings justify the performance gain of NTC with respect to point-to-point R-D function at lower rates. However, looking at higher rates, we find no evidence of binning occurring in the source space (not depicted). This is also consistent with the R-D performances achieved by this model at high rates, more clearly visible in Fig.~\ref{fig:x=y+n_var_n=0.1}, which are significantly worse than the WZ R-D bound. We attribute the noticeable suboptimal performance of NTC at high rates to its inability to learn sufficiently steep and high frequency functions in order to execute any efficient and flexible binning schemes, such as the periodic ones depicted in Fig.~\ref{fig:visualizations_gaussian}. Note that this observation is coherent with the analysis detailed in Section~\ref{sec:ntc_intro}. Since our marginal formulation operates on an unordered categorical latent space (see Section~\ref{subsec:estimating_bounds} for the relevant discussion), instead of an ordered one on real line as is the case for NTC, it is easier for our proposed compressor to recover more versatile binning mechanisms than those that could be achieved by the NTC-based compressor. This is also reflected in the improved R-D performances achieved by our marginal formulation provided in all panels of Fig.~\ref{fig:RD_results_gaussian}. We defer comparison with other related works~\cite{NDSC, ecsq_w_side_info} and our previous work~\cite{ozyilkan2023learned} to Section~\ref{sec:discussion}.

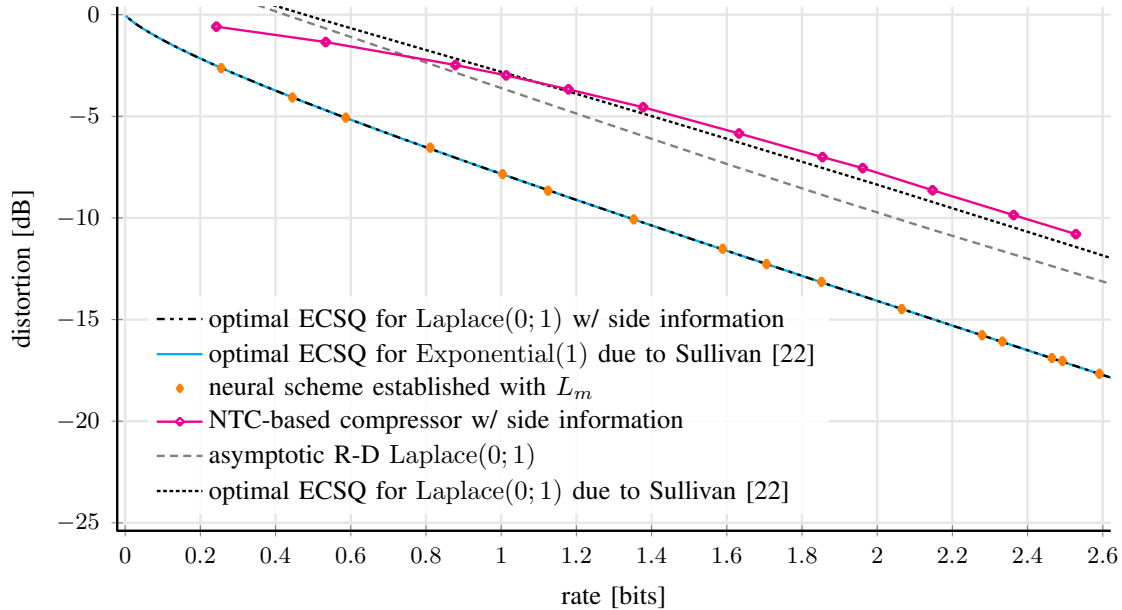
\begin{figure*}[t]
    \centering
    \begin{tikzpicture}[trim axis right]
    \begin{axis}[
      height=.30\textheight,
      width=.8\linewidth,
      scale only axis,
      xlabel={rate [bits]},
      ylabel={distortion [dB]},
      xmin=0.,
      xmax=2.6,
      ymin=-25.,
      ymax=0.,
      legend pos=south west,
      legend style={font=\small},
      reverse legend,
      ]
       \addplot[color=black, densely dotted] table {
      2.74308 -12.688400043360188
      2.56751 -11.655000043360188
      2.00991 -8.417500043360187
      1.50291 -5.556480043360187
      1.26059 -4.222570043360188
      1.05281 -3.095010043360188
      0.76978 -1.575450043360188
      0.51619 -0.21139004336018807
      0.250851 1.2708499566398122
      0.106253 2.163196956639812
      0.0551177 2.521273956639812
      0.00238992 2.977434956639812
      };
       \addplot[color=gray,densely dashed] table {
       2.74308 -13.915600043360186
       2.56751 -12.947800043360187
       2.13267 -10.499400043360188
       1.8955 -9.117600043360188
       1.59269 -7.297800043360187
       1.26059 -5.243130043360187
       1.05281 -3.940480043360188
       0.820686 -2.482050043360188
       0.632445 -1.300720043360188
       0.449202 -0.14804004336018783
       0.337394 0.563029956639812
       0.270387 0.9960199566398122
      };
      \addplot[color=magenta, mark=halfsquare*] table{
      0.2426915 -0.5877767130732536
      0.53348315 -1.3451550900936127
      0.8783622 -2.474716752767563
      1.0134085 -2.9860273003578186
      1.1787941 -3.6717602610588074
      1.3775315 -4.553080797195435
      1.6324406 -5.842599272727966
      1.8546665 -7.005162239074707
      1.9615476 -7.54610538482666
       2.1473196 -8.637049198150635
       2.3627374 -9.857673048973083
       2.528376 -10.79639196395874
      };
     \addplot[color=orange, mark=diamond*, only marks] table{
        0.25530246 -2.62782140986
        0.44477543 -4.072385132312775
        0.58686894 -5.06421252505
        0.8113903 -6.540983319282532
        1.0037888 -7.846092581748962
        1.124822 -8.65842982546
        1.3526752 -10.0701491477
       1.5890312 -11.5160469653
        1.7056725 -12.2720961215
        1.8517447 -13.1422869803
        2.0652654 -14.4811932685
        2.2790368 -15.7666353823
        2.3329997 -16.0783259513
      2.4644983 -16.8913733959198
      2.4927237 -17.03464150428772
    2.5910053 -17.665698528289795
      };
        \addplot[color=cyan] table {
        2.7742027930105997 -18.785274783850056
        1.9968246064495052 -14.059477010209605
        1.515411082139869 -11.093318831718413
        1.1770338248299443 -8.968760154307365
        0.9248340090286805 -7.346774483304259
        0.7309059225521122 -6.063032646142112
        0.5791536438400673 -5.024449745589495
        0.4592349040357366 -4.172502040049563
        0.3639795120626874 -3.467498912574212
        0.28814109566804225 -2.880928884872328
        0.22773403540136747 -2.391378957669089
        0.1796528896370207 -1.9822023028459168
        0.1414383079876411 -1.6401087653143276
        0.11112348127838338 -1.3542720666872214
        0.08712729304083434 -1.1157403055998345
        0.06817605103829918 -0.9170315326904719
        0.053243736271954153 -0.7518460510137284
        0.041504978885045715 -0.6148545278545412
        0.03229728782946271 -0.5015367192135871
        0.02509034126919169 -0.40805491911261704
        0.010533546063756546 -0.19889235770449876
        0.005856352988695899 -0.12130092798325709
        0.0017806358848948879 -0.04364218379077078
      };
        \addplot[dashdotdotted] table {
        2.7742027930105997 -18.785274783850056
        1.9968246064495052 -14.059477010209605
        1.515411082139869 -11.093318831718413
        1.1770338248299443 -8.968760154307365
        0.9248340090286805 -7.346774483304259
        0.7309059225521122 -6.063032646142112
        0.5791536438400673 -5.024449745589495
        0.4592349040357366 -4.172502040049563
        0.3639795120626874 -3.467498912574212
        0.28814109566804225 -2.880928884872328
        0.22773403540136747 -2.391378957669089
        0.1796528896370207 -1.9822023028459168
        0.1414383079876411 -1.6401087653143276
        0.11112348127838338 -1.3542720666872214
        0.08712729304083434 -1.1157403055998345
        0.06817605103829918 -0.9170315326904719
        0.053243736271954153 -0.7518460510137284
        0.041504978885045715 -0.6148545278545412
        0.03229728782946271 -0.5015367192135871
        0.02509034126919169 -0.40805491911261704
        0.010533546063756546 -0.19889235770449876
        0.005856352988695899 -0.12130092798325709
        0.0017806358848948879 -0.04364218379077078
      };
      \legend{optimal ECSQ for $\mathrm{Laplace}(0;1)$ due to Sullivan~\cite{sullivan}, asymptotic R-D $\mathrm{Laplace}(0;1)$, NTC-based compressor w/ side information, neural scheme established with $L_m$, optimal ECSQ for $\mathrm{Exponential}(1)$ due to Sullivan~\cite{sullivan}, optimal ECSQ for $\mathrm{Laplace}(0;1)$ w/ side information};
    \end{axis}
    \end{tikzpicture}
     \caption{Rate--distortion performances (R-D) obtained with marginal formulation as in Eq.~\eqref{eq:L_m} and NTC-based compressor with side information that uses a variant of Eq.~\eqref{eq:classic_rd} as the objective function. As in Fig.~\ref{fig:ntc_initial}, we consider a simple one-shot source coding with side information setting, where $X \sim \mathrm{Laplace}(0;1)$  and $Y= \mathrm{sgn}(X)$, i.e., the sign function of the input realization. The performances of the optimal ECSQ coincide with either the (point-to-point) entropy--distortion function (Definition~\ref{def:E_D}) or the entropy--distortion function with side information (Definition~\ref{def:entropy_opt_side_info}). Note that the optimal ECSQ for the Laplacian source having sign information available at the decoder corresponds to the optimal ECSQ for the exponential source.}
     \label{fig:RD_results_laplacian}
\end{figure*}

\begin{figure*}[t]
\centering
\begin{subfigure}{.5\textwidth}
  \centering
  \includegraphics[width=0.95\linewidth]{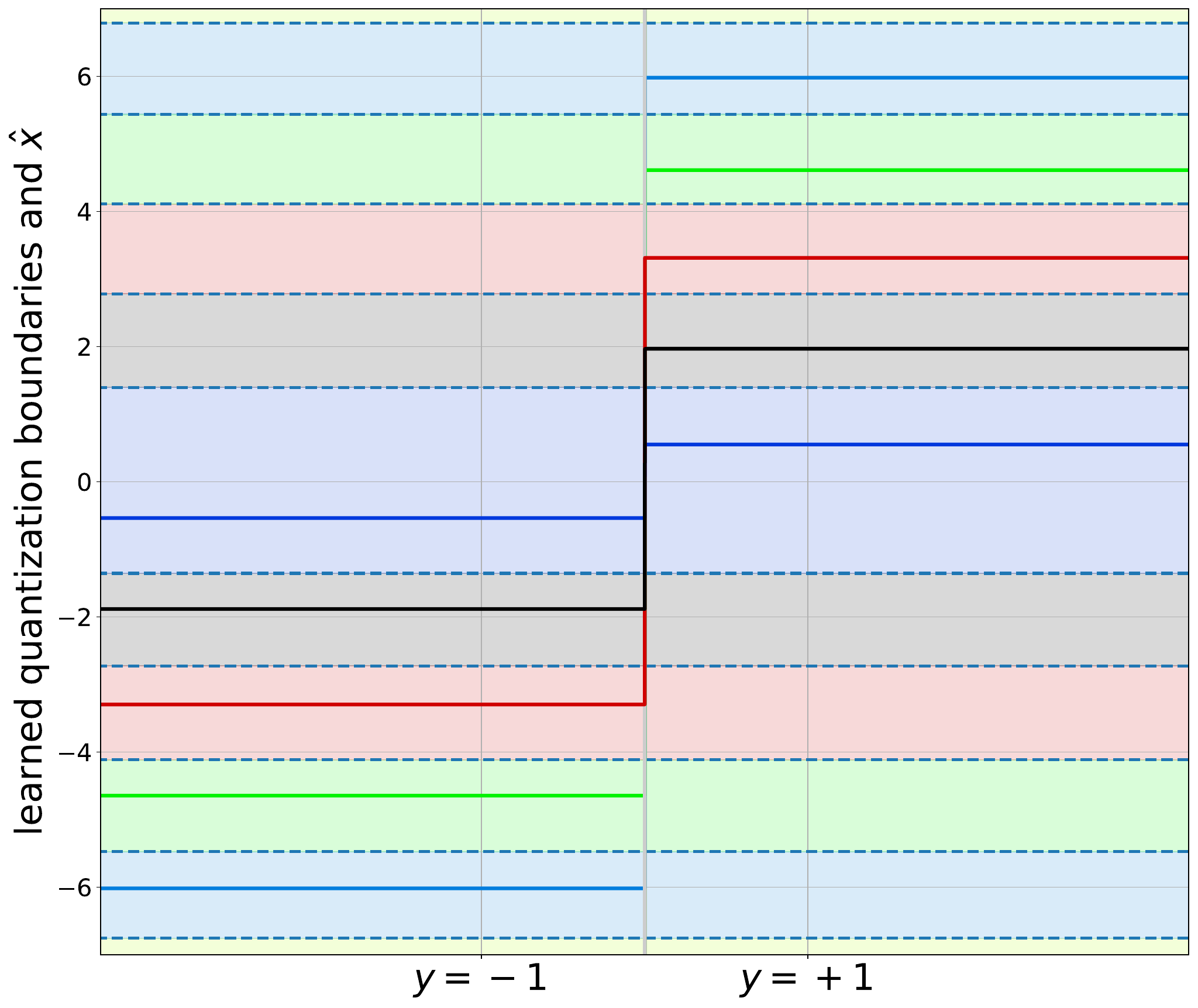}
  \label{fig:visualization_laplacian_1}
\end{subfigure}%
\hfill
\begin{subfigure}{.5\textwidth}
  \centering
  \includegraphics[width=0.95\linewidth]{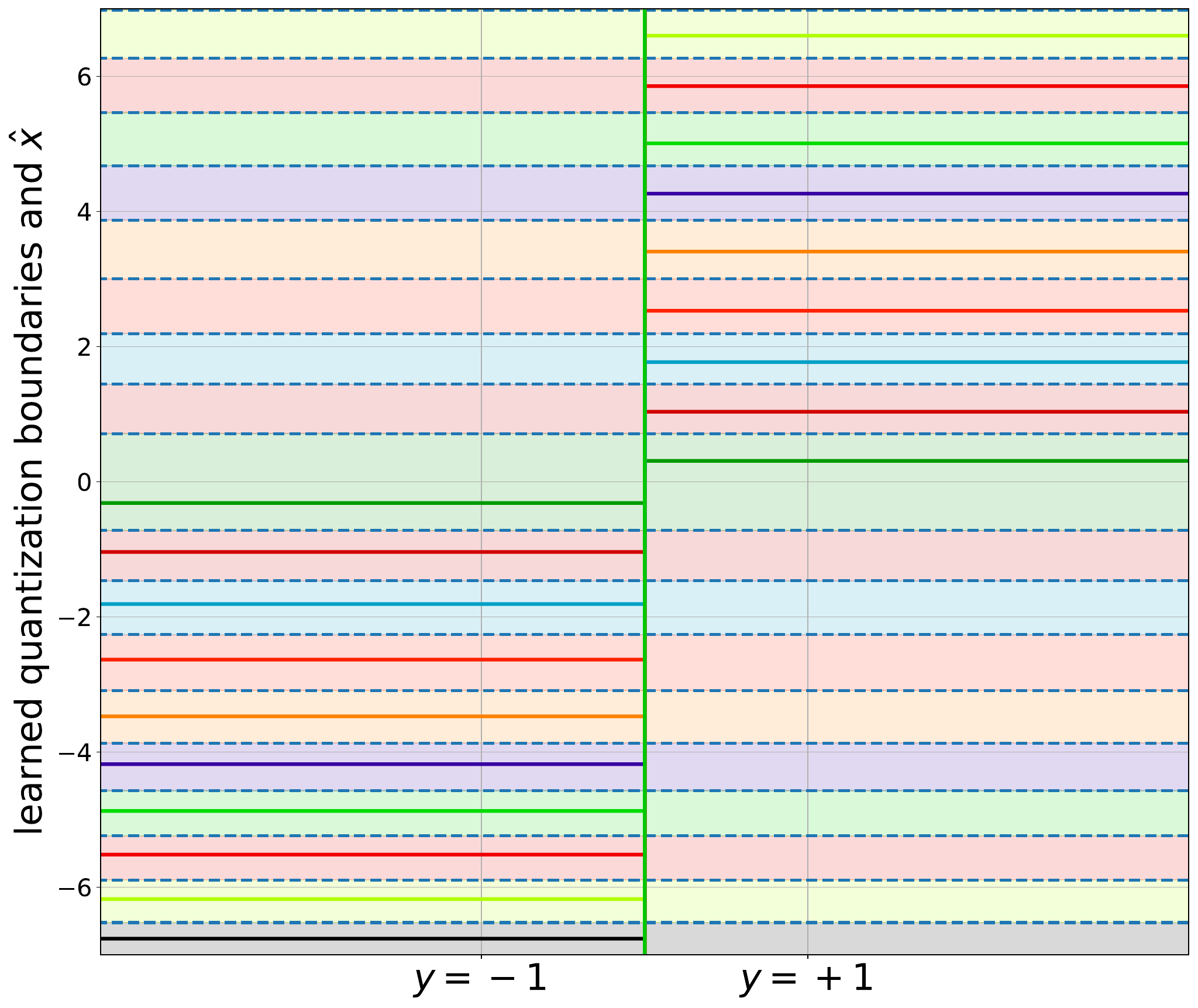}
  \label{fig:visualization_laplacian_2}
\end{subfigure}
\caption{Visualizations (best viewed in color) of the learned encoder $u =  \argmin_{k} \ell_{\mathrm{m}}(k, x)$ (see Eq.~\eqref{eq:f_m}) and neural decoder $\hat{x} = g_{\boldsymbol{\phi}}(u,y)$ of the marginal formulation (see Eq.~\eqref{eq:L_m}), for the Laplacian experiment described in Section~\ref{sec:experimental_setup}. The dashed horizontal lines are quantization boundaries, and the colors between boundaries represent unique values of $u$. We depict the decoding function as separate plots for each value of $u$, using the same color assignment. The visualized models on the left and right panels achieve $-8.41$ dB at $1.10$ bits and $-13.53$ dB at $1.92$ bits, respectively. Note that both of the visualized models achieve same
optimal entropy--distortion with side information performance as the optimal entropy-constrained scalar quantizer for $\mathrm{Exponential}(1)$, characterized by Sullivan in~\cite{sullivan}.}
\label{fig:visualizations_laplacian}
\end{figure*}

\subsection{Laplacian Experiments}
\label{subsec:laplacian}

Next, we consider a setup to evaluate how close our formulation can get the the theoretical optimum in the one shot-case, which we used to establish the suboptimality of NTC framework in Section~\ref{sec:ntc_intro}, that involves a Laplacian source. Note that nature of the correlation structure in Laplacian experiments is different than that of the Gaussian setup. In the latter case, the side information is only statistically correlated with the input source whereas in the former one, it is directly a deterministic function of the input realization.

Since the compression performance that could be achieved by the optimum ECSQ design is already characterized by Sullivan~\cite{sullivan} for exponential distributions (see Section~\ref{sec:experimental_setup} for the relevant discussion), unlike the comparisons in Section~\ref{subsec:quadratic_Gaussian}, we now plot the empirical results versus one-shot operational R-D function (see Definition~\ref{def:entropy_opt_side_info}) of the optimal ECSQs due to Sullivan. The plot in Fig.~\ref{fig:RD_results_laplacian} shows that our marginal formulation recovers the theoretically optimum entropy--distortion function with side information. Looking at the visualizations obtained by this model, provided in both panels of Fig.~\ref{fig:visualizations_laplacian}, we remark that it learns symmetric mappings in the source space, whose quantization boundaries are visibly sensitive to the R-D region where the model operates in. Because the side information in this case is the sign function of the input realization, assigning the quantization indices in a symmetric fashion with respect to $x=0$ corresponds to the optimal grouping strategy. The figure demonstrates that our model exhibits this desired behavior.

\subsection{Discussion}
\label{subsec:discussion}

We have proposed two practice-oriented solutions to the WZ problem, whose optimal theoretical solution is asymptotic and non-constructive. By establishing two variants of neural upper bounds to the entropy--distortion function with side information, we have introduced constructive learning based-compressors, operating in the one-shot quantization regime in tandem with variable rate entropy coding. Explicitly visualizing the behavior of these models, we provided post-hoc interpretations for the learned encoders and decoders. To ensure that the learning procedure cannot benefit from prior knowledge of the source, imposed into the design via special structure, we opted for a very generic parametrization for the models. Figs.~\ref{fig:visualizations_gaussian} and~\ref{fig:visualizations_laplacian}, along with the results in our previous work~\cite{ozyilkan2023learned}, provide the first explicit evidence of ANN-based learned compressors recovering some elements of the optimal theoretical solution, both through binning with respect to the source space, and piecewise linear behavior of the decoding function for the quadratic-Gaussian case, and achieving the theoretical optimum in the one-shot quantization case for the Laplacian experiments. Binning is a heavily used mathematical tool in information theory, and also characterizes practice-oriented schemes such as DISCUS~\cite{DISCUS}. Unlike the systematic partitioning of the quantized source space through cosets, as in DISCUS, our models are data-driven, and may find practical use for other sources beyond the Gaussian or Laplacian cases, most of whose feasible solutions are unknown to this day. Our findings provide interesting data-driven insights about the nature of a classical source coding problem with side information.

In terms of constructive solutions, we have established the link between two neural upper bounds (Section~\ref{subsec:estimating_bounds}) and two corresponding operational schemes (Section~\ref{subsec:operational_schemes}), by picking a suitable entropy coding technique for each one. In the case of the marginal formulation in Eq.~\eqref{eq:L_m}, it is attainable with high-order classic entropy coding, operating on discrete values. Considering the conditional formulation in Eq.~\eqref{eq:L_c}, we make use of an ideal SW coding scheme~\cite{Slepian:IT:73}, which compresses sufficiently large blocks of quantized source elements to the rate of $H(f_{c}(X) \vert Y)$. Unlike in the marginal variant, the operational role of SW coding is to additionally exploit the correlation between $f_{c}(X)$ and $Y$ to yield further compression. This explains our empirical finding that in this case, there is no binning observed in the quantization (as SW coding takes care of this). State-of-the-art channel coding schemes such as LDPC~\cite{swc_nq, tcq_ldpc, ldpc_0, ldpc_3,  ldpc_4, girod_rate_adaptive_ldpc, girod_image_authentication} and turbo codes~\cite{turbo_0, turbo_3, turbo_4, turbo_1, turbo_2, blum_SW} have been demonstrated to yield results coming close to the theoretical SW bound. To be fair, in order to achieve optimality, these schemes make certain assumptions about the virtual channel, which might not be met in our case.

Compared to our previous methods in~\cite{ozyilkan2023learned}, we established simpler and more robust learning-based solutions to the WZ problem that do not rely on various hyperparameter choices. These were mainly due to the Concrete distribution formulation~\cite{concrete} (e.g., a temperature parameter dictating the amount of relaxation from the discrete distribution) as a means of learning unordered categorical latent spaces at the encoder output. 
Instead, we now immediately minimize upper bounds as in Eqs.~\eqref{eq:L_m} and~\eqref{eq:L_c}, which directly corresponds to the discretized setup used at test time. Our encoding functions, as in Eqs.~\eqref{eq:f_m} and~\eqref{eq:f_c}, are also already set to be deterministic from the beginning of the training, and we do not require any sampling from the encoders as in~\cite{ozyilkan2023learned}. One additional system requirement that our new formulations have, in comparison to previous work, is the availability of conditional sampling, in the form of $y \sim p(y \vert x)$, at the encoder side (see Eqs.~\eqref{eq:l_m} and~\eqref{eq:l_c}).

Tu \emph{et al.}~\cite{ecsq_w_side_info} studied the ECSQ designs for the WZ problem, where they specifically assumed Gaussian sources as inputs (see Section~\ref{sec:relevant_quantizer_designs} for the relevant discussion). Similar to our conditional formulation (see Fig.~\ref{fig:sys_conditional}), their scheme has a scalar quantizer in tandem with a SWC. Their results, only reported for $Y=X+N$ with $X$ and $N$ being independent Gaussian, are replicated in Fig.~\ref{fig:y=x+n_var_n=0.1}. We remark that our conditional formulation, also coupled with an ideal SWC, achieves a performance similar to or better than the scheme in~\cite{ecsq_w_side_info}. As an added bonus, our formulation does not make any prior assumptions on the distribution of information sources.

Previous work in~\cite{NDSC} investigated the construction of neural WZ schemes using a machine learning technique called VQ-VAE~\cite{vq_vae}, which is comparable to both of our formulations in that its objective is amenable to optimization using SGD. However, the objective in~\cite{NDSC} does not explicitly model entropy, and instead contains a proxy objective that encourages utilization of all elements available in the codebook. It thus does not correspond to an optimization of the R-D Lagrangian (as in Eqs.~\eqref{eq:l_m} and \ref{eq:l_c}). The results of \cite{NDSC}, obtained for $Y=X+N$ with $X$ and $N$ being independent Gaussian, are reproduced in Fig.~\ref{fig:y=x+n_var_n=0.01}. We note that both of our methods as well as the NTC-based compressor with side information outperform this scheme. We attribute the suboptimal performance of the scheme to the lack of explicit accounting for entropy in the learning objective.

Notable prior work on the machine learning side~\cite{VIB,CEB} is related to the \emph{information bottleneck} problem~\cite{tishby}. While their learning objectives are quite similar to our marginal and conditional formulation, respectively, the authors are strictly concerned with probabilistic model fitting and/or \emph{representation learning}. Hence, they neither explore connections to operational compression schemes, nor do they consider the effects of quantization and/or binning.

In their present form, our operational schemes inherit the limitations of VQ, namely that increasing the codebook size naively is computationally infeasible. A topic for future work would be to investigate extensions that enable compression of high-dimensional sources such as images. In the case of SW coding, learned channel coding techniques~\cite{kim_1, kim_2} could be explored to relax the assumptions about the virtual channel arising between the quantized source and side information.

\appendix

\begin{figure}[H]
    \centering
    \includegraphics[width=0.6\linewidth]{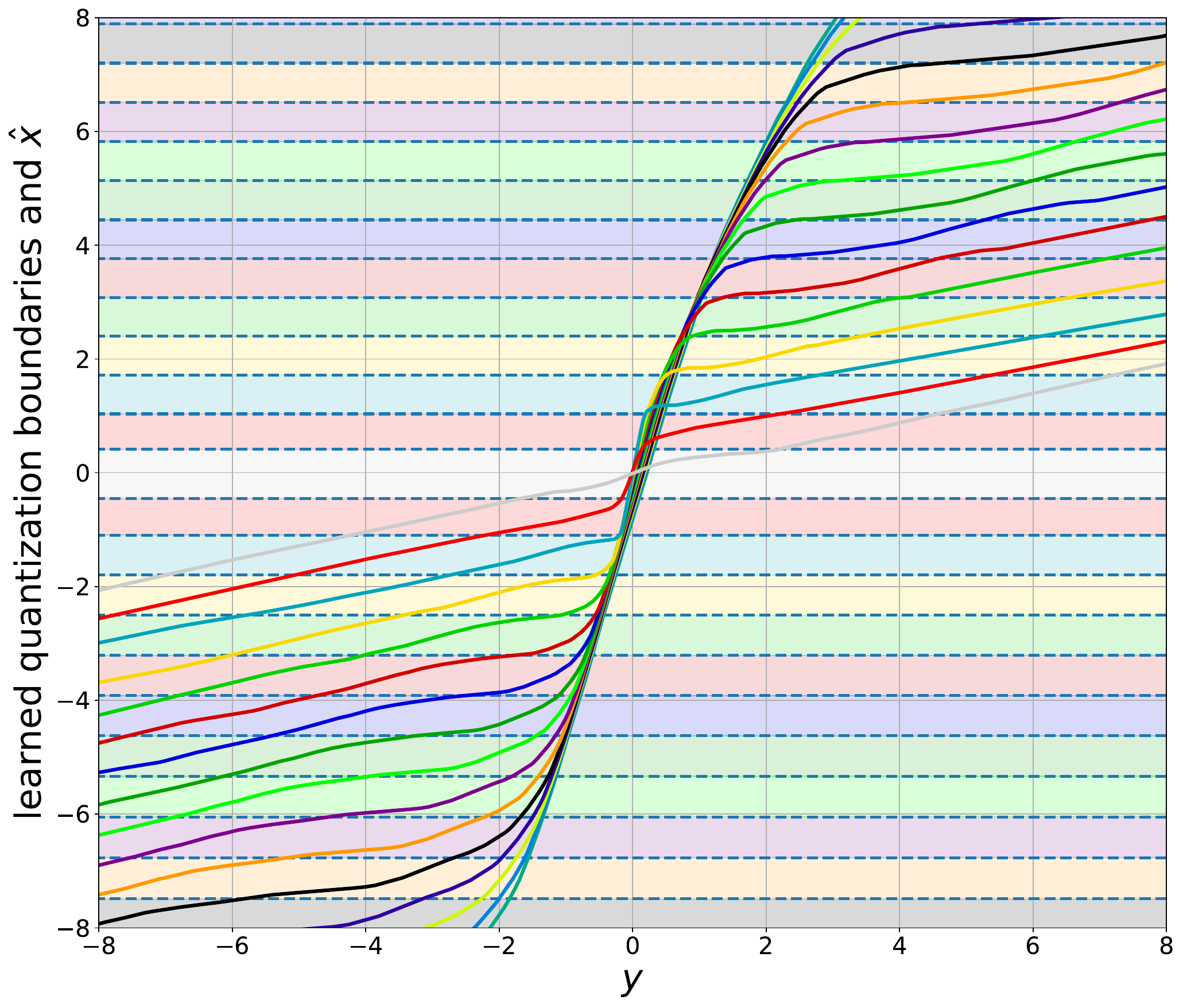}
    \caption{Visualization (best viewed in color) of an NTC-based compressor with side information, having a neural encoder $u=\nint{f_{\boldsymbol{\theta}}(x)}$ and a neural decoder $\hat{x} = g_{\boldsymbol{\phi}}(u,y)$ (see Eq.~\eqref{eq:classic_rd}), for the Gaussian WZ setup with the correlation structure of $X=Y+N$ having independent $Y \sim \mathrm{N}(0,1)$ and $N \sim \mathrm{N}(0,10^{-1})$ (as in Fig.~\ref{fig:x=y+n_var_n=0.1}). The dashed horizontal lines are quantization boundaries, and the colors between boundaries represent unique values of $u$. We depict the decoding function as separate plots for each value of $u$, using the same color assignment. The visualized model achieves $-13.96$ dB at $2.07$ bits, yielding better rate--distortion (R-D) performance than the point-to-point R-D function in Eq.~\eqref{eq:rd_p2p}.}
    \label{fig:NTC_vis}
\end{figure}

\bibliographystyle{IEEEtran}

\bibliography{ref.bib}

\end{document}